\documentclass[fleqn,usenatbib]{mnras}

\usepackage{newtxtext,newtxmath}

\usepackage[T1]{fontenc}

\DeclareRobustCommand{\VAN}[3]{#2}
\let\VANthebibliography\thebibliography
\def\thebibliography{\DeclareRobustCommand{\VAN}[3]{##3}\VANthebibliography}

\usepackage{graphicx}	
\usepackage{amsmath}	
\usepackage[flushleft]{threeparttable}
\usepackage{hyperref}

\newcommand{\ha}{H$\mathrm{\alpha}$}   
\newcommand{\hb}{H$\mathrm{\beta}$}  
\newcommand{\oiii}{[\mbox{O\,\textsc{iii}}]}
\newcommand{\nii}{[\mbox{N\,\textsc{ii}}]}
\newcommand{\sii}{[\mbox{S\,\textsc{ii}}]}
\newcommand{\oii}{[\mbox{O\,\textsc{ii}}]}
\newcommand{\oi}{[\mbox{O\,\textsc{i}}]}
\newcommand{\siii}{[\mbox{S\,\textsc{iii}}]}
\newcommand{\hii}{\mbox{H\,\textsc{ii}}}
\newcommand{\ncl}{$N_{\mathrm{cluster}}$}
\newcommand{\mcl}{$M_{\mathrm{cl}}$}
\newcommand{\msol}{M$_{\odot}$}
\newcommand{\hdenini}{$n_{\mathrm{H,0}}$}

\newcommand{\noise}{Noise-Net}
\newcommand{\noises}{Noise-Nets}
\newcommand{\normal}{Normal-Net}
\newcommand{\normals}{Normal-Nets}
\defcitealias{Kang+22}{Paper 1}

\title[Noise-Net]{Noise-Net: Determining physical properties of \hii\ regions reflecting observational uncertainties}

\author[D. E. Kang et al.]{
Da Eun Kang,$^{1}$\thanks{E-mail: kang@uni-heidelberg.de (DK)} 
Ralf~S.~Klessen,$^{1,3}$ 
Victor~F.~Ksoll,$^{1}$ 
Lynton~Ardizzone,$^{2}$ 
Ullrich Koethe$^{2}$  \newauthor
and Simon~C.~O.~Glover$^{1}$
\\
$^{1}$Universit\"{a}t Heidelberg, Zentrum f\"{u}r Astronomie, Institut f\"{u}r Theoretische Astrophysik, Albert-Ueberle-Stra{\ss}e 2, D-69120 Heidelberg, Germany\\
$^{2}$Universit\"{a}t Heidelberg, IWR, Computer Vision and Learning Lab, Berliner Stra{\ss}e 43, D-69121 Heidelberg, Germany\\
$^{3}$Universit\"{a}t Heidelberg, Interdisziplin\"{a}res Zentrum f\"{u}r Wissenschaftliches Rechnen, Im Neuenheimer Feld 205, D-69120 Heidelberg, Germany\\ 
}

\date{Accepted XXX. Received YYY; in original form ZZZ}

\pubyear{2023}

\begin{document}
\label{firstpage}
\pagerange{\pageref{firstpage}--\pageref{lastpage}}
\maketitle

\begin{abstract}
Stellar feedback, the energetic interaction between young stars and their birthplace, plays an important role in the star formation history of the universe and the evolution of the interstellar medium (ISM). Correctly interpreting the observations of star-forming regions is essential to understand stellar feedback, but it is a non-trivial task due to the complexity of the feedback processes and degeneracy in observations.
In our recent paper, we introduced a conditional invertible neural network (cINN) that predicts seven physical properties of star-forming regions from the luminosity of 12 optical emission lines as a novel method to analyze degenerate observations. We demonstrated that our network, trained on synthetic star-forming region models produced by the WARPFIELD-Emission predictor (WARPFIELD-EMP), could predict physical properties accurately and precisely.
In this paper, we present a new updated version of the cINN that takes into account the observational uncertainties during network training. Our new network named \noise\ reflects the influence of the uncertainty on the parameter prediction by using both emission-line luminosity and corresponding uncertainties as the necessary input information of the network. We examine the performance of the \noise\ as a function of the uncertainty and compare it with the previous version of the cINN, which does not learn uncertainties during the training. We confirm that the \noise\ outperforms the previous network for the typical observational uncertainty range and maintains high accuracy even when subject to large uncertainties.

\end{abstract}




\section{Introduction}
Newly-formed stars and  Giant Molecular Clouds (GMCs), the stellar birthplace, interact with each other via stellar feedback. Young massive stars born in a GMC inject a large amount of energy and momentum into the surrounding environment via stellar winds, radiation, thermal pressure from photoionized gas or supernovae~\citep{Krumholz+14, Klessen&Glover16}. The feedback from young stars can disrupt further star formation by destroying the GMC or can locally promote new star formation~\citep{Shetty&Ostriker08, Dale+13, Rahner+19, Chevance+20, Kim+21, Grudic+22}.

To understand the influence of stellar feedback and the physical properties of the star-forming region on star formation, it is essential to correctly interpret the observed star-forming regions. However, the overall physical process occurring in the star-forming region is very complicated because the multiple feedback mechanisms are non-linearly coupled together and act simultaneously. Moreover, observations of star-forming regions are usually degenerate which means that different physical systems look similar in the observational space. Observational uncertainties as well as the degeneracy and complex physics make it even more difficult to describe the observation with theoretical models by using a classical fitting method.

We apply artificial neural networks~\citep[NNs;][]{Goodfellow+16} to solve this complicated problem in star-forming regions. NNs can connect physical parameters and observational measurements through a statistical model, without designating a specific physical model. Recently, various machine learning techniques including NNs have been utilized in many astronomical fields e.g., to classify observations~\citep{Wu+19, Walmsley+21, Whitmore+21}, to identify structures or exoplanets~\citep{Abraham+18, DeBeurs2022}, and to predict physical parameters~\citep{Fabbro+18, Ksoll+20, Olney+20, Sharma+20, Shen+22}. 
In this study, we adopt the supervised learning approach, a type of machine learning that trains the network using labelled data sets, as we want to estimate specific physical parameters we want from quantities we can measure from observed star-forming regions.

In degenerate systems, the forward process that translates the physical parameters to observations is usually well-defined but involves information loss, from which the degeneracy arises. Due to the lost information, it is difficult to solve the inverse problem using a classical neural network trained through the inverse process. Therefore, we need a neural network that provides a full posterior probability distribution of the physical system conditioned on the given observation.

A conditional invertible neural network~\citep[cINN;][]{Ardizzone+19b, Ardizzone+21}, a type of invertible neural network~\citep[INN;][]{Ardizzone+19a}, is a deep learning architecture specialized for the inverse inference of degenerate systems. Unlike classical neural networks, a cINN is trained through the forward process of the system but also learns about the inverse process for free due to its invertibility. By using the additional latent variables in the network, the cINN captures the information otherwise lost during the forward process training and provides a full posterior distribution of physical parameters through the inverse process.

In \citet[][hereafter \citetalias{Kang+22}]{Kang+22}, we introduced a cINN that predicts seven physical parameters of \hii\ regions using 12 optical emission-line luminosities. In this first study, we trained our network using synthetic \hii\ region models produced by WARPFIELD-Emission predictor~\citep[WARPFIELD-EMP;][]{Pellegrini+20}, the pipeline modelling the synthetic observation based on the 1-dimensional stellar feedback code WARPFIELD~\citep{Rahner+17,Rahner+18}.
We confirmed that our network predicts each parameter very accurately and precisely and validated the network predictions by re-simulating the emission-line luminosity of predicted models.
Although the cINN presented in \citetalias{Kang+22} did not consider observational error in its prediction by definition, we introduced a way to obtain the posterior distribution reflecting the observational errors by modifying the posterior sampling method. Large observational errors worsened the overall performance, but we confirmed that our network was accurate in most cases unless the smallest error among 12 lines was larger than 0.1\%.

In this study, we introduce a new type of cINN, which we term \noise,  that always reflects observational uncertainties in the prediction. Unlike our first cINN in \citetalias{Kang+22}, \noise\ learns not only about the relationship between physical parameters and luminosities, but also about the influence of luminosity errors during the network training. In this paper, we focus on comparing the performance of \noise\ with the type of cINN used in \citetalias{Kang+22}.

The paper is structured as follows. In Section~\ref{section:methodology}, we explain the overall methodology used in this paper, including the structure and training of the cINNs and synthetic \hii\ region database. In Section~\ref{section:result}, we compare the performance of two cINNs as a function of the observational error using a statistical approach and individual models. We discuss the implication of our main results on the physical aspects and machine learning aspects in Section~\ref{section:discussion} and summarise the results of this paper in Section~\ref{section:summary}.

\section{Methodology}
\label{section:methodology}
\subsection{cINN and Normal-Net}
\label{subsection:cinn}

The cINN architecture~\citep{Ardizzone+19b, Ardizzone+21} is an inverse problem solver specialized in degenerate systems. We assume that, in a degenerate system, different sets of physical parameters (\textbf{x}) can be mapped onto an identical observation (\textbf{y}) because of the information loss during the forward process. To capture the lost information, the cINN introduces the third component, the latent variables (\textbf{z}), and builds a bijective mapping between \textbf{x} and \textbf{z}. The observation \textbf{y} is applied as a condition \textbf{c} to this mapping in both the forward ($f$) and inverse process:
\begin{equation} 
\label{eq:cINN-basic}
\begin{split}
    \mathbf{z} = f(\mathbf{x}; \mathbf{c} = \mathbf{y}), \\
    \mathbf{x} = g(\mathbf{z}; \mathbf{c} = \mathbf{y}),
\end{split}
\end{equation}
where $g=f^{-1}$ \citep{Ardizzone+21}. Due to the bijective mapping, the dimension of \textbf{z} is the same as the dimension of \textbf{x}.

We train the network through the forward process to learn $f$ and prescribe the latent variables to follow a standard multivariate normal probability distribution $p(\textbf{z}) = N(0, \textbf{I})$ with zero mean and unit covariance matrix, where \textbf{I} is the identity matrix with a dimension of $\mathrm{dim}(\textbf{z}) \times \mathrm{dim}(\textbf{z})$. As the cINN consists of invertible affine coupling blocks, the cINN automatically learns the inverse process $g$ from the forward training. Following the inverse process $g$ and sampling the latent variables from the prescribed distribution $p(\mathbf{z})$, we can obtain the posterior distribution of \textbf{x} for the given condition (\textbf{c}), $p(\mathbf{x}|\mathbf{c}=\mathbf{y})$.

In \citetalias{Kang+22}, we introduced a cINN that predicts seven physical parameters from the luminosity of 12 optical emission lines. The seven parameters are initial cloud mass \mcl, star formation efficiency $\epsilon$, cloud density \hdenini, age of the first generation cluster (i.e., the oldest cluster), age of the youngest cluster, the number of clusters and the evolutionary phase of the cloud. 
The network presented in \citetalias{Kang+22}, which we will refer to as \textbf{\normal} in the following, used a basic cINN structure and training method predicting physical parameters from the given observations without considering the observational errors, $p(\mathbf{x}|\mathbf{y})$. Although the \normal\ does not learn about observational errors ($\boldsymbol{\sigma}$) during the training, in \citetalias{Kang+22}, we introduced a method to account for the observational uncertainties, $p(\mathbf{x}|\mathbf{y}, \boldsymbol{\sigma})$, through a modification of the posterior sampling procedure and analysed the performance of the network as a function of the observational error (see Section 7 of \citetalias{Kang+22}).

In this paper, we introduce \textbf{\noise}, a variant of the cINN architecture and training method that can predict posterior distributions accounting for the observation errors, $p(\mathbf{x}|\mathbf{y}, \boldsymbol{\sigma})$, without the additional steps required for the \normal\ by learning the influence of errors during the training.
In the following (Section~\ref{subsection:noise_training}), we describe the structure and training procedure of the \noise\ in order to predict the parameters considering both observations and corresponding errors. 

\subsection{\noise\ and noise training}
\label{subsection:noise_training}

\noise\ and its training method are based on SoftFlow introduced by \cite{Kim+20}.
\cite{Kim+20} used the following ideas in their training to improve the performance of the network that reconstructs a two-dimensional pattern such as a spiral line from the latent variables prescribed to an isotropic 2D Gaussian distribution. Firstly, \cite{Kim+20} randomly sampled the error (i.e., 1-$\sigma$ width of the Gaussian distribution) and perturbed the original pattern (X) by adding the Gaussian noise based on the sampled error. Then they trained the network with the perturbed data (X') and use the sampled error as a condition. \cite{Kim+20} found that the pattern restoration accuracy of the network depended on the amount of error given by the condition. The network successfully reconstructed a clean pattern given the small error, and this result was better than those of the other networks trained without error. 
The situation in our study is slightly different to \cite{Kim+20} in that we use the cINN that already has a condition \textbf{y} and that we want to deal with the error of \textbf{y}, not the error of \textbf{x}. However, the idea of SoftFlow allowed us to design the \noise\ that takes into account observation errors during the training.

We want the \noise\ to predict physical parameters considering the given observation (luminosity) and the uncertainty of the observations (i.e., luminosity error). Thus, the \noise, by definition, uses both \textbf{y} and the error of \textbf{y}, $\boldsymbol{\sigma}$, as a condition of the network: $\textbf{c} = [\textbf{y}, \boldsymbol{\sigma}]$. In this study, we define the luminosity error $\boldsymbol{\sigma}$ as a fractional 1-sigma unitless uncertainty which is normalized by the corresponding luminosity value. Because of the additional condition, the dimension of \textbf{c} in the \noise\ is twice as large as in the \normal. As we use the information on 12 emission lines, the dimension of \textbf{c} of the \noise\ in this paper is 24.

For the \noise\ to learn the influence of the observational uncertainties, introducing the error as an additional condition of the network alone is not enough. We also need to modify the training strategy of the network. In this paper, we refer to the method of training \noises\ as noise training. The differences between noise training and normal training are two additional steps processed during the training on the fly.

The first step is to randomly sample the luminosity errors for each training model. For the network to learn about various $\boldsymbol{\sigma}$ values, the $\boldsymbol{\sigma}$ is not included in the training data but is randomly sampled from a given distribution at every training epoch and for every training model during the training. 
In this paper, we sample the errors in a logarithmic scale because the range of error values we want to train is wide. 
For each training model, the luminosity error of the $i-$th emission line, $\sigma_{i}$, is sampled from the uniform distribution, 
\begin{equation} 
\label{eq:sample_sigma}
    p(\mathrm{log}\:\sigma_{i}) = U(a, b).
\end{equation}
We sample the luminosity error of all 12 emission lines from one probability distribution (Eq.~\ref{eq:sample_sigma}) with the same minimum (a) and maximum value (b) to simplify the training setup but it is also possible to use a different probability distribution for each emission line. In this study, we use the minimum error of -5 and maximum error of -0.5 which are equivalent to 0.001 and 31.6\%\ error respectively.

The second step is to perturb the luminosity of the training model by adding random Gaussian noise to the true luminosity values ($\mathbf{y^*}$) based on the $\boldsymbol{\sigma}$ sampled in the first step. The perturbed luminosity of the $i-$th emission line ($y_{i}'$) is calculated by
\begin{equation} 
\label{eq:perturb_lum}
  y_{i}' = y^{*}_{i} \: (1 + r_{i}), \mathrm{where} \; r_{i} \in N(0, \sigma^{2}_{i}).
\end{equation}
In order to prevent the perturbed luminosity value from being negative because of a large amount of noise, we clip the perturbed luminosity to the minimum value of 1.

After these two steps, we train the network by using the true parameter values ($\mathbf{x^*}$) as an input and the group of perturbed luminosity and randomly sampled luminosity error ([$\mathbf{y'}$, $\boldsymbol{\sigma}$]) as a condition to the forward process. 
In \citetalias{Kang+22}, we trained the network (\normal) by using the true parameters values and true luminosity values ($\mathbf{x^*}$ and $\mathbf{y^*}$) as an input and a condition (i.e., normal training method), so that the \normal\ learned about the same values for each training epoch repeatably during the training. On the other hand, during the noise training, the \noise\ learns about various luminosity and error values from an identical training model because errors and perturbation noises are randomly sampled at every training epoch. 
Consequently, the prediction power of the trained \noise\ varies as a function of the luminosity error.

\subsection{Training data: WARPFIELD-EMP}
\label{subsection:emp}
To train the cINN, which adopts a supervised learning approach, we need numerous sets of data containing both the observable quantities (\textbf{y}) and physical parameters that we want to predict (\textbf{x}). However, it is difficult to collect a sufficient number of well-analyzed \hii\ regions from real observations. Hence, to train and test the networks, we use the same database used in our first paper (\citetalias{Kang+22}), which consists of 505,748 synthetic \hii\ region models that we produced by using the WARPFIELD-EMP pipeline \citep{Pellegrini+20}. 
Based on the evolution of a massive star-forming cloud described by the 1D stellar feedback code WARPFIELD~\citep{Rahner+17, Rahner+18}, WARPFIELD-EMP produces the continuum and line emission of the model cloud at a large series of output times by using CLOUDY~\citep{Ferland+17} to calculate the continuum and line emissivities and POLARIS~\citep{Reissl+16} to compute the transfer of the resulting radiation through the cloud. 
In this section, we briefly summarise our synthetic \hii\ region models and database that is described in detail in \citetalias{Kang+22}.

\subsubsection{Synthetic \hii\ regions}

WARPFIELD~\citep{Rahner+17}, which is the basis of our synthetic model, simulates the evolution of an isolated massive cloud acted on by stellar feedback. At the beginning of the evolution ($t=0$), a star cluster with mass $M_{*}$ is formed at the centre of the cloud. The cluster mass is determined by two initial conditions, the initial cloud mass (\mcl) and the star formation efficiency ($\epsilon$): $M_{*} = \epsilon M_{\mathrm{cl}}$. WARPFIELD treats star formation in a highly idealized way and assumes that all stars in the cluster form instantaneously.
Due to the stellar feedback produced by the cluster, the cloud is separated into distinct zones: a diffuse central bubble (i.e., inner free wind zone), a dense shell surrounding the bubble with hot shocked wind materials, and a static cloud region outside of the shell. The evolution of the cloud is described by solving the equation of motion of the dense shell considering several feedback mechanisms such as stellar winds, supernovae, thermal gas pressure, radiation pressure, and gravity. WARPFIELD assumes that within the shell the gas is in quasi-hydrostatic equilibrium, and that the evolution of the cloud can be distinguished into four distinct evolutionary phases according to the dynamics of the shell: Phase 1, 2, 3, and 0.

In the first evolutionary phase, Phase 1, the shell expands rapidly, driven by the thermal pressure of the hot shocked wind material. During this phase, the evolution of this hot gas is adiabatic and the influence of gravity and radiation pressure can be ignored. Phase 1 ends once the inner bubble loses its hot gas, either because the hot gas radiatively cools (which occurs after a time $t_{\text{cool}}$) or because the bubble bursts and the hot gas leaks out. In WARPFIELD~\citep{Rahner+17}, it is assumed for simplification that the bubble bursts only when the shell sweeps up the entire natal cloud ($t_{\text{sweep}}$, i.e., when $R_{\text{shell}} > R_{\text{cloud, initial}}$). If the bubble cools down before the shell sweeps up the whole materials in the cloud ($t_{\text{cool}} < t_{\text{sweep}}$), the evolutionary phase turns into Phase 2, otherwise it turns into Phase 3 directly. After Phase 1, the shell expansion is dominated by radiation pressure and the ongoing ram pressure exerted by supernovae and stellar winds. The counteracting effects of the gravity of the star cluster and the self-gravity of the shell are now no longer negligible.

The fate of the clouds is divided into two options, depending on the balance between stellar feedback and gravity. If the feedback is more dominant than gravity, the shell continues to expand into the low-density interstellar medium after sweeping up the whole material in the cloud (Phase 3). The density of the shell gradually decreases and the shell finally dissolves into the ambient ISM. The evolution ends when the maximum number density of the shell is smaller than 1 cm$^{-3}$ for more than 1~Myr.
On the other hand, if gravity becomes more dominant than the feedback, the shell stops expanding and recollapses to the centre. This collapse triggers the formation of a new star cluster when the shell radius becomes smaller than 1~pc. The recollapsing phase until the birth of the new cluster is labelled as Phase 0. For simplicity, it is assumed that the star formation efficiency of the new birth is the same as the first burst ($M_{*,2} = \epsilon (M_{\text{cl}} - M_{*,1}) $). The evolution continues with the new shell formed by the stellar feedback dominated by the new cluster.

In the WARPFIELD model, stars within the cluster follow a Kroupa initial mass function~\citep{Kroupa01} and the time-dependent effect of the stellar feedback from the evolving star cluster is calculated with STARBURST99~\citep{Leitherer+99, Leitherer+14} by using the Geneva evolution tracks~\citep{Ekstrom+12, Georgy+12, Georgy+13}. Hence, the spectral energy distribution (SED) of the WARPFIELD model is time-dependent due to the evolving cluster and cloud and especially complex if the SED of multiple clusters with different ages are combined within one cloud due to the recollapse.

Based on the time-dependent evolution information of the WARPFIELD model, WARPFIELD-EMP uses CLOUDY~\citep[C17,][]{Ferland+17}, a spectral synthesis code, to calculate emissivities of various lines and continuum processes. We first run CLOUDY for the dense shell to obtain the emissivities as a function of radial position within the shell. If the natal cloud surrounding the shell still remains, we run CLOUDY a second time to calculate the emissivities from the static cloud by using the output of the first CLOUDY run for the shell as an incident flux.
Lastly, the radial information on the emissivities is passed on to POLARIS~\citep{Reissl+16, Reissl+19} to calculate the final luminosity information considering the attenuation within the shell and the natal cloud. In WARPFIELD-EMP, POLARIS is used in ray-tracing mode to solve the radiative transfer equation for the rays passing through a 3D grid of the shell and cloud. We use a 3D spherical grid for POLARIS because the WARPFIELD model is a 1D spherical symmetric model. 
Finally, the output of POLARIS is projected onto a 2D space and the 2D map is spatially integrated to obtain the velocity-integrated 1D luminosity of lines and continuum of a given WARPFIELD model at a certain time.

Our synthetic model contains information on many fundamental physical parameters together with the corresponding continuum emission and a series of lines for a given WARPFIELD model at a given time. We only use seven parameters and 12 emission lines in this study, but WARPFIELD-EMP originally provides many other lines within a wide frequency range from optical to radio listed in Table D1 and D2 in \cite{Pellegrini+20}.

\subsubsection{Training set and test set}
In our first study (\citetalias{Kang+22}), we built and introduced a new database that consists of 505,748 WARPFIELD-EMP synthetic \hii\ region models evolved from 10,000 initial WARPFIELD clouds. In this study, we use the same database to train and evaluate our networks.

A WARPFIELD-EMP model in our database is uniquely determined by four independent physical parameters: initial mass of the WARPFIELD cloud (\mcl), star formation efficiency (SFE), initial cloud density (\hdenini), and the age of the cloud ($t$). The first three parameters are the initial conditions of the WARPFIELD calculation. As the evolution of the WARPFIELD cloud begins, the age becomes the fourth independent parameter, which is equivalent to the age of the first cluster, because the evolution of WARPFIELD starts with the formation of the first cluster.
There are several other parameters that can be varied in the initial conditions of a WARPFIELD cloud (e.g., metallicity), but we fixed all of these parameters at constant values. In particular, we use solar metallicity and a constant radial profile of the initial density and turn off the effects of turbulence and the magnetic field for all models in our database.

For the initial 10,000 WARPFIELD clouds, we randomly sampled \mcl, SFE, and \hdenini\ of the clouds within the range of 10$^{5-7}$~\msol, 2--10~\%, and 100--500~cm$^{-3}$, respectively. We evolve the clouds until they reach an age of 30~Myr, although depending on the physical conditions of the cloud, the evolution can end earlier if the shell dissolves into the ambient ISM.
WARPFIELD stores the time-dependent evolution in the prescribed time interval and we adopt a 0.1~Myr interval for our clouds. However, for computational efficiency reasons, we do not calculate the emission for all saved models, but instead run CLOUDY and POLARIS only when the physical conditions of the cloud (e.g., shell density, shell mass, shell radius, and ionizing photon flux) change by a large enough amount to cause a significant change in the emission or when the evolutionary phase of the cloud changes. Therefore, the time interval of the database is not constant due to the adoptive time sampling method. In particular, the time intervals tend to grow longer as the cloud ages, because the emission changes more rapidly in the early evolutionary phase. For more details of the time distribution of the outputs, we refer the reader to Figure 1 in \citetalias{Kang+22}.

Although the initial cloud parameters are uniformly distributed, the physical parameters of the models in the database are not evenly distributed (see both Figure~\ref{fig:sample} in this work and Figure 1 in \citetalias{Kang+22}), because each cloud evolves differently and terminates its evolution at different ages. For example, if the gravity of the cloud is more dominant than the stellar feedback and the cloud recollapses repeatedly, this cloud survives longer and saves more models than clouds with strong stellar feedback. Hence, the distribution of \mcl\ and SFE are biased toward a large mass and small SFE, respectively. Additionally, the proportion of models with only one cluster (single cluster model) and the proportion of models in Phase 3 are much higher than the other cases.

The distribution of training data affects network performance. In the previous study, we demonstrated that the network trained with this database performs better for \hii\ regions with more common characteristics in the database. Specifically, the posterior distribution was usually more accurate and narrower for young and single cluster models rather than old models with multiple clusters. For this reason, it is better to sample the training data as evenly as possible or to make the distributions even through post-processing. However, it is also not easy, because the evolution of a cloud is a very complex function of the four independent parameters, so that it is difficult to predict the evolution from the given initial conditions. If we additionally sample more clouds with less populated characteristics in the current distributions, this could in turn introduce new biases in the other parameters. Therefore, we use the current database without additional modification, although the current distribution is not entirely optimal in terms of training.

We randomly divide the database with 505,748 models into a training set and a test set, using the former to train the network, while the latter serves to evaluate the trained network. In this study, we use 90\%\ of the database (455,174 models) to train the network and use the remaining 10\%\ (50,574 models) to evaluate the performance of the networks.

Our database contains various physical parameters and luminosity of several lines and continuum for each WARPFIELD-EMP model but we only use 12 optical emission lines and seven physical parameters in our study (Table~\ref{table:param_obs}), which are the same choices as in \citetalias{Kang+22}. The seven target parameters consist of initial cloud mass (\mcl), star formation efficiency (SFE), initial cloud density (\hdenini), age of the first cluster ($t$), age of the youngest cluster ($t_{\text{youngest}}$), the number of the clusters (\ncl), and the evolutionary phase of the cloud. We choose 10 optical lines within the range of 3700--9000~\AA: [\mbox{O\,\textsc{ii}}] 3726\AA,
[\mbox{O\,\textsc{ii}}] 3729\AA,  \hb\ 4861\AA, [\mbox{O\,\textsc{iii}}] 5007\AA, [\mbox{O\,\textsc{i}}] 6300\AA, \ha\ 6563\AA, [\mbox{N\,\textsc{ii}}] 6583\AA, [\mbox{S\,\textsc{ii}}] 6716\AA,  [\mbox{S\,\textsc{ii}}] 6731\AA, and [\mbox{S\,\textsc{iii}}] 9531\AA. We also add the total strength of the \oii\ and \sii\ doublets, referred to as [\mbox{O\,\textsc{ii}}] 3727\AA\ (blend) and [\mbox{S\,\textsc{ii}}] 6720\AA\ (blend).

\renewcommand{\arraystretch}{1.25}
\begin{table}
    \centering
    \caption{List of the seven physical parameters (i.e., \textbf{x} of the network) and list of the twelve emission lines whose luminosities are used as \textbf{y}. The information listed in this table is the same as Tables 1 and 2 in \citetalias{Kang+22}.}
    \label{table:param_obs}
    \begin{tabular}{l r}
        \hline
        \hline
        Parameter & Symbol \\
        \hline
        initial mass of star-forming cloud &  \mcl\ [M$_{\odot}$] \\
        initial star formation efficiency & SFE [\%] \\
        initial cloud number density &  \hdenini\ [cm$^{-3}$] \\
        age of the first cluster & $t$ [Myr]	\\
        age of the youngest cluster & $t_{\mathrm{youngest}}$ [Myr] \\
        number of star clusters & \ncl\ \\
        evolutionary phase of the cloud & Phase \\
        \hline
        \hline

        \hline
        \hline
        Line & Wavelength \\
        \hline
        \oii\ &  3726\AA  \\
        \oii\ (blend) & 3727\AA \\
        \oii\ &  3729\AA  \\
        \hb\ & 4861\AA \\
        \oiii\ & 5007\AA \\
        \oi\ & 6300\AA \\
        \ha\ & 6563\AA \\
        \nii\ & 6583\AA \\
        \sii\ & 6716\AA \\
        \sii\ (blend) & 6720\AA \\
        \sii\ & 6731\AA \\
        \siii\ & 9531\AA \\
        \hline
        \hline
    \end{tabular}
\end{table}

\subsection{Network setup} 
\label{subsection:network_setup}
The goal of this study is to introduce the new \noise\ together with the noise training and to compare the performance of the \noise\ and \normal. Thus we train one \normal\ and one \noise\ using the same network setup except for the intrinsic differences between the \normal\ and the \noise\ described in the previous section (e.g., the dimension of \textbf{c}). In this section, we explain the network architecture of our two networks and the data pre-processing steps required to train or use the network. Most of the setup used in this study is the same as in \citetalias{Kang+22}, except for some hyperparameters.

\subsubsection{Network construction} 

To construct the cINN architecture, we use the FrEIA~\citep[Framework for Easily Invertible Architectures;][]{Ardizzone+19a, Ardizzone+21} which is based on the `PyTorch' library~\citep{Paszke+19} as in \citetalias{Kang+22}.

The cINN consists of a series of affine coupling blocks that follow the architecture proposed by \cite{Dinh+16}. In this study, we use 16 affine coupling blocks to build a network, two times more than that of the network in \citetalias{Kang+22}. 
Each affine coupling block splits the input \textbf{u} into two parts (i.e., $\mathbf{u_1}$ and $\mathbf{u_2}$) and passes each part through affine transformations following
\begin{equation} 
\label{eq:forward affine}
\begin{split}
\mathbf{v_{1}} &= \mathbf{u_{1}} \odot \mathrm{exp}(s_{2}(\mathbf{u_{2}}, \mathbf{c})) + t_{2}(\mathbf{u_{2}}, \mathbf{c}), \\
\mathbf{v_{2}} &= \mathbf{u_{2}} \odot \mathrm{exp}(s_{1}(\mathbf{v_{1}}, \mathbf{c})) + t_{1}(\mathbf{v_{1}}, \mathbf{c}).
\end{split}
\end{equation}
The outputs of the two affine transformations (i.e., $\mathbf{v_1}$ and $\mathbf{v_2}$) are coupled into the final output \textbf{v}.

The invertibility of the cINN architecture comes from the invertibility of each affine coupling block. After training the cINN through the forward process, we can use the inverse process of the trained network following 
\begin{equation} 
\label{eq:inverse affine}
\begin{split}
\mathbf{u_{2}} &= (\mathbf{v_{2}} - t_{1}(\mathbf{v_{1}}, \mathbf{c})) \odot \mathrm{exp}(-s_{1}(\mathbf{v_{1}}, \mathbf{c})), \\
\mathbf{u_{1}} &= (\mathbf{v_{1}} - t_{2}(\mathbf{u_{2}}, \mathbf{c})) \odot \mathrm{exp}(-s_{2}(\mathbf{u_{2}}, \mathbf{c})).
\end{split}
\end{equation}

The internal transformations $s_{i}$ and $t_{i}$ do not need to be invertible themselves, because they are only evaluated in the forward direction in both the forward and inverse process of the cINN. In this paper, we adopt a single sub-network as an internal transformation of each affine coupling block~\citep[i.e., the GLOW configuration;][]{Kingma&Dhariwal18}. For each sub-network, we apply a simple fully connected architecture with 6 layers and a width of 256, using the rectified linear units (ReLU) as the activation functions.
After each affine coupling block, we add an invertible permutation layer to mix the information stream. The permutation layer is a random orthogonal matrix and is fixed during the training.

As shown in Eq.~\ref{eq:forward affine} and \ref{eq:inverse affine}, the condition (\textbf{c}) of the cINN architecture is always used as an additional input for each transformation in both the forward and inverse processes. However, before applying the condition to the affine coupling blocks, we pass the condition through an additional feed-forward network, the conditioning network, to extract higher-level features. According to \cite{Ardizzone+19b}, using a conditioning network in complex systems helps the efficient conditioning of the cINN by reducing the burden of the main network (i.e., a series of invertible blocks) to re-learn higher-level features in each affine coupling block.
The conditioning network can be either pretrained or trained together with the main network~\citep{Ardizzone+19b}. In this study, we jointly train the main network and the conditioning network because the latter method helps extract features that are more relevant to \textbf{x}. For the conditioning network, we adopt a simple fully connected feed-forward network with three layers and a width of 512.

Compared to \citetalias{Kang+22}, we double the number of affine coupling blocks and the number of layers of each internal sub-network, because deepening the network into the current setup improves the performance, especially for the \noise. In Appendix~\ref{section:deeper networks}, we will cover the influence of network depth on the prediction power of \noises\ and \normals\ in detail.

After constructing the network with the above setup, we train the network to minimize the maximum log-likelihood loss,
\begin{equation} 
\label{eq:loss}
\mathcal{L} = \mathbb{E}_{i} \Big[ \frac{\|f(\mathbf{x}_{i}; \mathbf{c}_{i}, \theta) \|^{2}_{2}}{2} - \mathrm{log}\:|J_i| \Big],
\end{equation}
as described in \cite{Ardizzone+19b} and \cite{Ksoll+20}, where $|J_i|$ denotes the determinant of the Jacobian matrix $|J_i| = \mathrm{det} \: \Big( \frac{\partial f}{\partial \mathbf{x}} |_{\mathbf{x}_i} \Big)$ and  $\mathbf{x}_{i}$ is the physical parameters of some training sample with the corresponding condition $\mathbf{c}_i$. 
During the training, we calculate the test loss, that is the loss calculated with the test set, as well as the training loss calculated with the training set. The network is trained until the deviation between the training loss and test loss is small and both losses converge.
The training time of the network depends on the batch size and the number of training epochs. Training one network for 100 epochs using a batch size of 256 took about 4 hours with NVIDIA GeForce RTX 2080 Ti graphic card and used about 1500 MB GPU memory and the size of the trained network is about 130MB. Training time and the size of the network also depend on the depth of the network (i.e., the number of affine coupling blocks and the number of layers of internal sub-network). By doubling both the number of layers and blocks compared to \citetalias{Kang+22}, training time and the network size have increased by about 2 and 4 times, respectively.

\subsubsection{Data pre-proccessing}

When training the cINN or using the trained cINN, we use pre-processed physical parameters, observations and observation errors as described in \citetalias{Kang+22}. 
For \ncl\ and phase that have discretized distributions with a sampling interval of 1, we smooth out the distributions by adding a small Gaussian noise with a standard deviation of 0.05. In \citetalias{Kang+22}, we demonstrated that smoothing out the discretized distribution improves the prediction power of the network (see Section 8.2. of \citetalias{Kang+22} for details). Please note that we smooth out these parameters only when we train the network.

For variables with a relatively broad range of values in linear space such as emission line luminosity (\textbf{y}), we convert them into logarithmic scale. In the case of the noise training, we convert the luminosity after perturbation ($\mathbf{y'}$ from Eq.~\ref{eq:perturb_lum}). As we sample the luminosity error $\boldsymbol{\sigma}$ from the wide distribution (Eq.~\ref{eq:sample_sigma}), we transform $\boldsymbol{\sigma}$ in log-scale as well. Next, we re-scale the distribution of \textbf{x}, \textbf{y}, and $\boldsymbol{\sigma}$ by using linear transformations. For physical parameters, $x_{i}$, we transform $x_{i}$ to $\hat{x}_{i}$ that has zero mean and unit standard deviation following
\begin{equation} 
\label{eq:x-rescale}
\hat{x}_{i} = (x_{i} - \mu_{x_{i}}) \cdot \frac{1}{s_{x_{i}}},
\end{equation}
where $\mu_{x_{i}}$ and $s_{x_{i}}$ are the mean and the standard deviation of the physical parameter $x_{i}$ calculated using the entire database. 
In the case of observation ($y_{i}$), we first centre them ($\tilde{y}_{i} = y_{i} - \mu_{y_{i}}$) and whiten the observation matrix following Equation 35 in \cite{Hyvarinen&Oja00} ($\mathbf{\hat{Y}} = \mathbf{W_{\tilde{Y}}} \mathbf{\tilde{Y}} $) to have distributions with unit variance for each emission line and identity matrix for the covariance matrix of observations. To calculate $\mu_{y_{i}}$ and $\mathbf{W_{\tilde{Y}}}$, we use true values ($\mathbf{y^{*}}$) of the entire database, not using the perturbed values, even for the \noise.

For the observation error $\sigma_{i}$, we apply the same linear transformation used for physical parameters following
\begin{equation} 
\label{eq:sigma-rescale}
\hat{\sigma}_{i} = (\sigma_{i} - \mu_{\sigma_{i}}) \cdot \frac{1}{s_{\sigma_{i}}}
\end{equation}
and change the mean and standard deviation of the distribution to zero and unity. As the errors are randomly sampled during the noise training from $p(\mathrm{log} \: \sigma)$ of Eq.~\ref{eq:sample_sigma} unlike the physical parameters and observations, we use the mean and the standard deviation of Eq.~\ref{eq:sample_sigma} for $\mu_{\sigma_{i}}$ and $s_{\sigma_{i}}$.
We transform $x_{i}$, $y_{i}$ and $\sigma_{i}$ to $\hat{x}_{i}$, $\hat{y}_{i}$, and $\hat{\sigma}_{i}$, respectively, when training the network. When using the inverse process to obtain a posterior distribution, we transform $y_{i}$ and $\sigma_{i}$ to $\hat{y}_{i}$, and $\hat{\sigma}_{i}$ and transform $\hat{x}_{i}$ calculated by the network to $x_{i}$.

\subsection{How to sample posterior estimates from the network}
\label{subsection:sampling}
As explained in Section~\ref{subsection:cinn}, the inverse process of the cINN calculates \textbf{x} from the corresponding condition \textbf{c} and latent variable \textbf{z}, following Eq.~\ref{eq:cINN-basic}. We obtain the posterior distribution $p(\mathbf{x}|\mathbf{c})$ by sampling the latent variables $N_{\mathrm{z}}$ times from the prescribed probability distribution $p(\textbf{z}) = N(0, \textbf{I})$ and then calculating the corresponding \textbf{x} using the inverse process $g$ in Eq~\ref{eq:cINN-basic}. Thus, the number of obtained posterior estimates is $N_{\mathrm{z}}$.
In this way, we obtain $p(\mathbf{x}|\mathbf{y})$ from the \normal\ and $p(\mathbf{x}|\mathbf{y}, \boldsymbol{\sigma})$ from the \noise. The standard \normal\ provides a posterior distribution without considering the error as it does not learn the error during the training. However, in \citetalias{Kang+22}, we introduced a Monte Carlo-based marginalisation method that allows us to obtain $p(\mathbf{x}|\mathbf{y}, \boldsymbol{\sigma})$ from the \normal. In this study, we only use the posterior distribution considering the observation error to compare the performance of the \noise\ and the \normal\ as a function of the error. Here, we summarise the marginalisation method introduced in \citetalias{Kang+22} to obtain posterior distributions that account for the errors from the \normal.

The overall process of the marginalisation method for the \normal\ is described by 
\begin{equation}
    \label{eq:marginalisation}
    p(\mathbf{x}|\mathbf{y},\boldsymbol{\sigma}) = \int p(\mathbf{x}|\mathbf{y'}) \: q(\mathbf{y'}|\mathbf{y}, \boldsymbol{\sigma}) \,d\mathbf{y'},
\end{equation}
where $q$ is a profile of a luminosity error, which is a Gaussian distribution in our paper, i.e., $N(\mathbf{y}, \boldsymbol{\sigma}^{2})$.
Given a set of 12 emission line luminosities (\textbf{y}) and the corresponding 1-sigma unitless errors ($\boldsymbol{\sigma}$) measured from observation, the first step is to perturb the luminosity in the same way that we do in the noise training, i.e.~following Eq~\ref{eq:perturb_lum}. In both cases, we assume that the perturbed luminosity follows a Gaussian distribution centred on the value $y$ with a standard deviation of $\sigma$.
By sampling Gaussian noise for each emission line $N_{\mathrm{p}}$ times, we obtain $N_{\mathrm{p}}$ sets of perturbed luminosity ($\textbf{y}'_{1}, ..., \textbf{y}'_{N_{\mathrm{p}}} $) from a given observation. Then, for the $i-$th perturbed luminosity set $\textbf{y}'_{i}$, we sample the latent variables $N_{\mathrm{z,normal}}$ times to get a posterior distribution $p(\textbf{x}|\textbf{y}'_{i})$. By superimposing the posterior distributions of all perturbed luminosity sets, we finally obtain the $p(\mathbf{x}|\mathbf{y}, \boldsymbol{\sigma})$ that consists of $N_{\mathrm{p}} \times N_{\mathrm{z,normal}}$ posterior samples from the \normal.

From \citetalias{Kang+22}, we found that it is important to produce a sufficient number of perturbed luminosity sets (i.e., $N_{\mathrm{p}}$) to get a smooth posterior distribution. In the previous study, we used $N_{\mathrm{p}}=3000$ and $N_{\mathrm{z,normal}}=100$, which are large enough to ensure a smooth distribution even when the $\boldsymbol{\sigma}$ is large. However, in the case of small $\boldsymbol{\sigma}$, a smaller $N_{\mathrm{p}}$ is enough to produce a smooth distribution because the range of the perturbation is narrow. To reduce the computation time required for the posterior sampling and analysis, we flexibly adjust $N_{\mathrm{p}}$ and $N_{\mathrm{z,normal}}$ values depending on the network performance and the magnitude of the error. Considering the range of error values used in this study, we keep $N_{\mathrm{z,normal}}$ fixed at 50 and vary $N_{\mathrm{p}}$ between 1000 and 2500 depending on the error. If the luminosity error is too large, the posterior sometimes shows a very spiky distribution despite large $N_{\mathrm{p}}$ and $N_{\mathrm{z,normal}}$. We confirm that in such cases, the spiky shape is not due to the lack of $N_{\mathrm{p}}$ or $N_{\mathrm{z,normal}}$ but an inevitable result of the large errors.
In the case of the \noise, we only need to decide the number of latent variable samples, $N_{\mathrm{z,noise}}$. In this study, we use $N_{\mathrm{z,noise}}$ of 20,000 when we use the \noise.


\section{Noise-Net vs. Normal-Net}
\label{section:result}
In this section, we compare the performance of the \normal\ and the \noise\ as a function of the observation error. As mentioned in Section~\ref{subsection:network_setup}, we trained both networks with the same setup except for the fundamental differences between the \normal\ and the \noise. We use the synthetic \hii\ region models of the test set instead of real observations to concentrate on the validation of the cINN rather than the validation of the synthetic training data.

\subsection{Experiment setup} 
\label{subsection:exp_setup}

\subsubsection{Sample selection}
\label{subsubsection:sample selection}

\begin{figure*}
    \includegraphics[height=0.87\columnwidth]{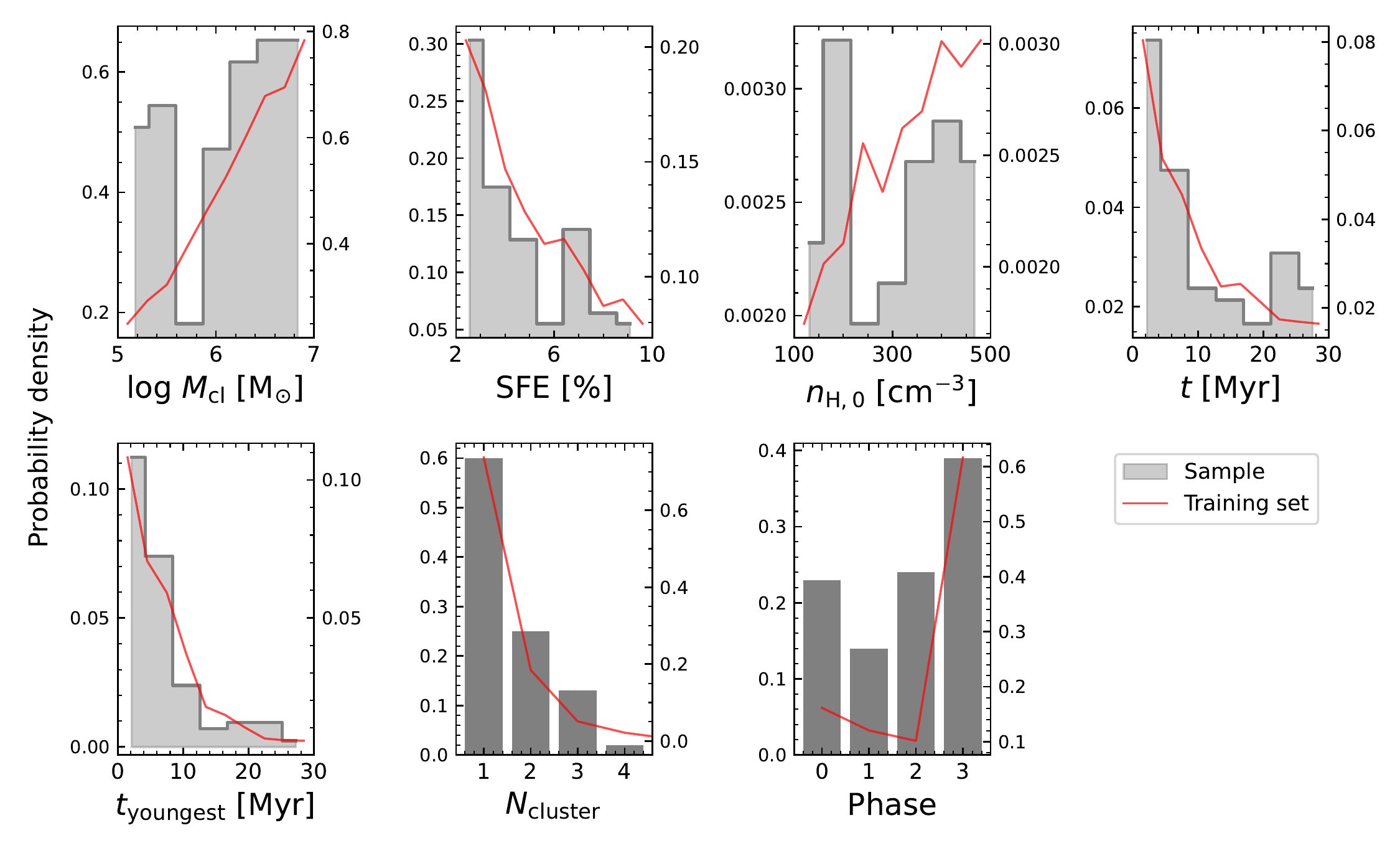}
	\includegraphics[height=0.87\columnwidth]{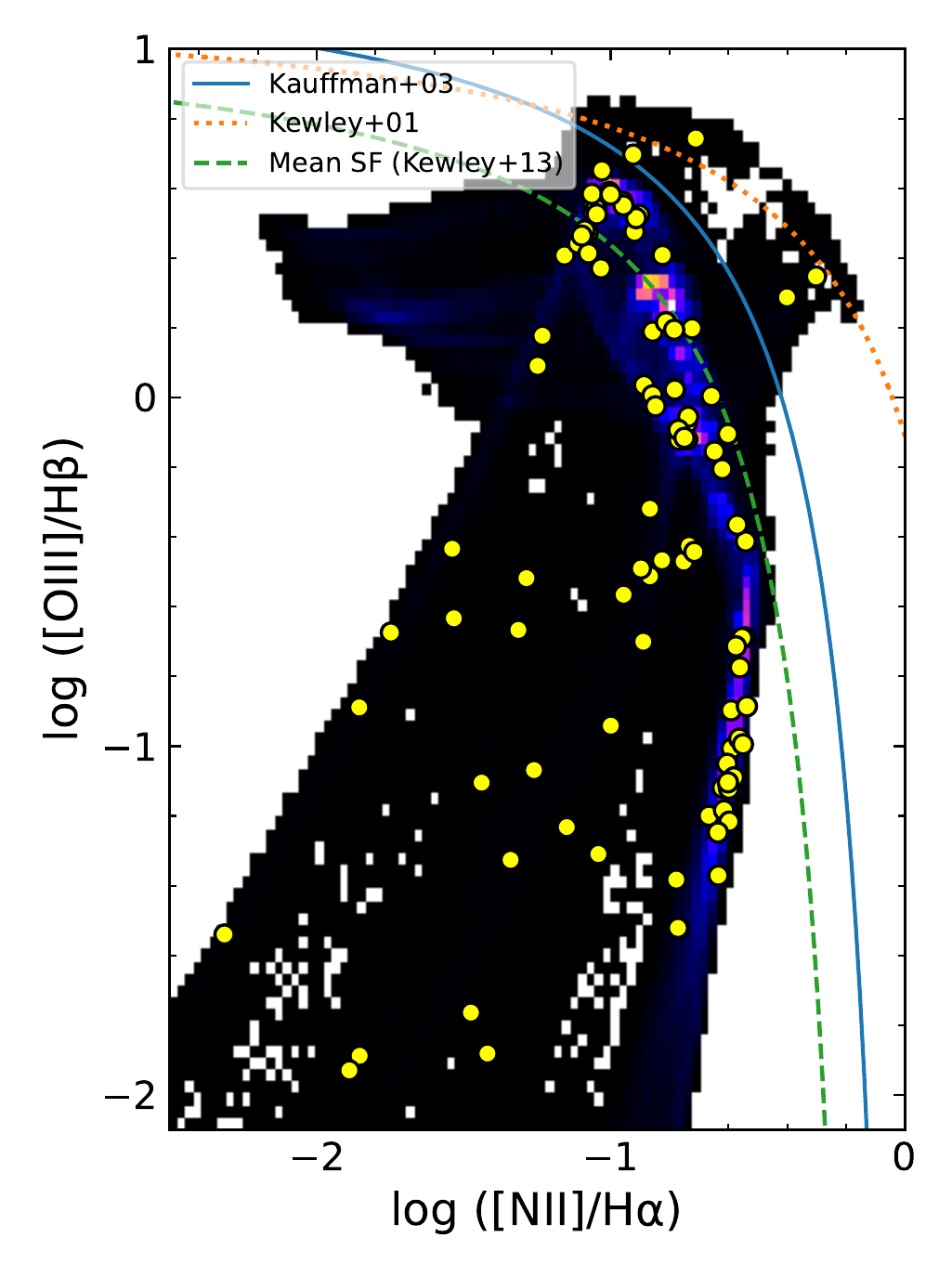}
    \caption{Distribution of seven physical parameters of the selected 100 test models (left) and their locations in the BPT diagram (right). In the left figure, the grey histogram and left y-axis of each panel show the distribution of the selected sample, and the red line and right y-axis show the distribution of 455,174 models of the training set. In the right figure, the background two-dimensional colour histogram shows the number density of the entire training set and yellow dots indicate the locations of 100 test models.
     } \label{fig:sample} 
\end{figure*}

To reduce the computation time for the evaluations, we only use 100 models among the 50,574 models in the test set as a representative subset. We randomly select these 100 test models based on the following criteria. Please note that we use the same selection criteria used in \citetalias{Kang+22} but select a new set of 100 test models. We present the parameter distributions of the selected models as well as their locations in the BPT diagram~\citep{Baldwin+81} in Figure~\ref{fig:sample}.

First, we exclude models with \nii/\ha\ < $10^{-3}$ or \oiii/\hb\ < $10^{-2}$, considering the typical line-ratio values of observed star-forming galaxies~\citep{Kauffmann+03, Kewley+06} and \hii\ regions~\citep{Sanchez+15, Rousseau-Nepton+18}. Next, we select 4 of the 100 models from the extreme area on the BPT diagram beyond the revised demarcation curve between active galactic nuclei (AGNs) and starburst galaxies proposed by \cite{Kauffmann+03}. As shown in Figure~\ref{fig:sample}, one of the four models is beyond the more extreme demarcation curve of \cite{Kewley+01}. For the other 96 models, we set a maximum fraction for \ncl\ and phase to prevent too many similar models from being selected. The distributions of the entire training data (red line in the left panel of Figure~\ref{fig:sample}) for these two parameters have biases toward the single cluster and Phase 3 models, respectively. We limit the fraction of single cluster models to 60\%\ and the fraction of Phase 3 models to 40\%\ to reduce this bias.

\subsubsection{Noise model and error selection}
\label{subsubsection:noise}
Our synthetic test models are, by definition, fully accurate without any uncertainty measurements. Thus, we assume a simple noise model, the same method used in \citetalias{Kang+22}, to assign mock luminosity errors for our experiments. Although diverse sources affect errors measured in real observations, we adopt a noise model that only considers Poisson noise. We assume that the error of each emission line is an independent random variable and ignore any covariance between physically related lines such as blended lines and their components. In our noise model, given the 1-sigma error of the brightest emission line, the errors of the remaining 11 emission lines are automatically determined following 
\begin{equation} 
    \label{eq:luminosity noise model}
    \sigma_{\mathrm{line}} = \sigma_{\mathrm{b}} \times \sqrt{\frac{L_{\mathrm{brightest}}}{L_{\mathrm{line}}}},
\end{equation}
where $\sigma_{\mathrm{b}}$ is the error of the brightest emission line of the observation, which is the smallest error among 12 emission lines.

Using the error of the brightest emission line ($\sigma_{\mathrm{b}}$) as the representative error of one observation, we investigate the influence of the error on the posterior distribution by increasing $\sigma_{\mathrm{b}}$. We select 16 values from 0.01\%\ to 10\%\ in 0.2 dex intervals on a logarithmic scale. For each network, we sample the posterior distributions of each parameter for the 100 test models with varying $\sigma_{\mathrm{b}}$. As mentioned in Section~\ref{subsection:sampling}, the number of posterior samples for one observation varies depending on the network and the size of the error. In Table~\ref{table:n_sample}, we list the numbers of sampling used in this experiment depending on the $\sigma_{\mathrm{b}}$ value.
Especially when the given error is large, our network sometimes returns physically incorrect or extremely extrapolated posterior estimates such as negative star formation efficiencies or cluster ages larger than 100~Myr. We exclude all of these unrealistic posterior estimates before our analysis.

\renewcommand{\arraystretch}{1.25}
\begin{table}
    \centering
    \caption{List of the numbers used to sample a posterior distribution depending on the error of the brightest emission line ($\sigma_{\mathrm{b}}$).
    \label{table:n_sample}}
    \begin{threeparttable}
    \begin{tabular}{p{0.15\textwidth}l   c   c   c }
        \hline
        \hline
        $\sigma_{\mathrm{b}}$ [\%] & $N_{\mathrm{p}}$ \tnote{1}  & $N_{\mathrm{z,normal}}$ \tnote{2}   & $N_{\mathrm{z,noise}}$ \tnote{3} \\
        \hline
         $\sigma_{\mathrm{b}}$ \: $\leq$ 0.1  &  1000 & 50 & 20000 \\
        0.1 $< \: \sigma_{\mathrm{b}} \:  \leq $ 1 & 2000 & 50 & 20000 \\
        1 $< \:  \sigma_{\mathrm{b}} \:  \leq $ 10 & 2500 & 50 & 20000 \\
        \hline
        \hline
    \end{tabular} 
    \begin{tablenotes}
      \small
      \item[1] the number of perturbed mock luminosity sets for the \normal\
      \item[2] the number of latent variable sampling for each mock luminosity set for the \normal\
      \item[3] the number of latent variable sampling for the \noise\
    \end{tablenotes}
    \end{threeparttable}
\end{table}

Although we use $\sigma_{\mathrm{b}}$ as the representative error of one observation, the errors of the other 11 emission lines of the 100 test models with the same $\sigma_{\mathrm{b}}$ value are all different, because these errors are determined by the luminosity ratio between the line and the brightest emission line, which is typically the \ha\ line. The error of the faintest emission line, usually the [\mbox{O\,\textsc{i}}] 6300\AA, is on average 17 times larger than the error of the brightest emission line. The error of the remaining 11 lines excluding the brightest emission line is on average 3.8 times larger than $\sigma_{\mathrm{b}}$.

\subsubsection{Evaluation methods}
\label{subsubsection:evaluation}
To evaluate the prediction power of the network, we introduce two evaluation indices used in this study that describe the accuracy and precision of the network prediction, respectively. 

The first index, which represents the accuracy of the network, is given by the deviation between the posterior estimate and the ground truth value of the test model. For \ncl\ and phase, we use the linear deviation ($X - X^{*}$) and for the other five parameters, we use the logarithmic deviation $\Big( \mathrm{log} \: \frac{X}{X^{*}} \Big) $ . 
To calculate the accuracy index we either use all posterior samples or only one representative estimate from the posterior distribution. For the representative value, we measure the maximum a posteriori (MAP) point estimates by performing a Gaussian kernel density estimation on the 1D posterior distribution of each parameter and finding the maximum of the derived probability density. 

The second index, the precision index, is the uncertainty interval at the 68\%\ confidence level (i.e., $u_{68}$). The $u_{68}$ value represents the width of the 1D posterior distribution similar to the width of $\pm 1$ standard deviation. 

For the 3200 posterior distributions obtained from the two networks for our 100 test models with 16 different $\sigma_{\mathrm{b}}$ values, we measure the accuracy and precision indices of each parameter. In the following sections, we compare the \normal\ and the \noise\ by using the measured performance indices.

\subsection{Statistical comparison} 
\label{subsection:statistical}

\begin{figure*}
	\includegraphics[width=2\columnwidth]{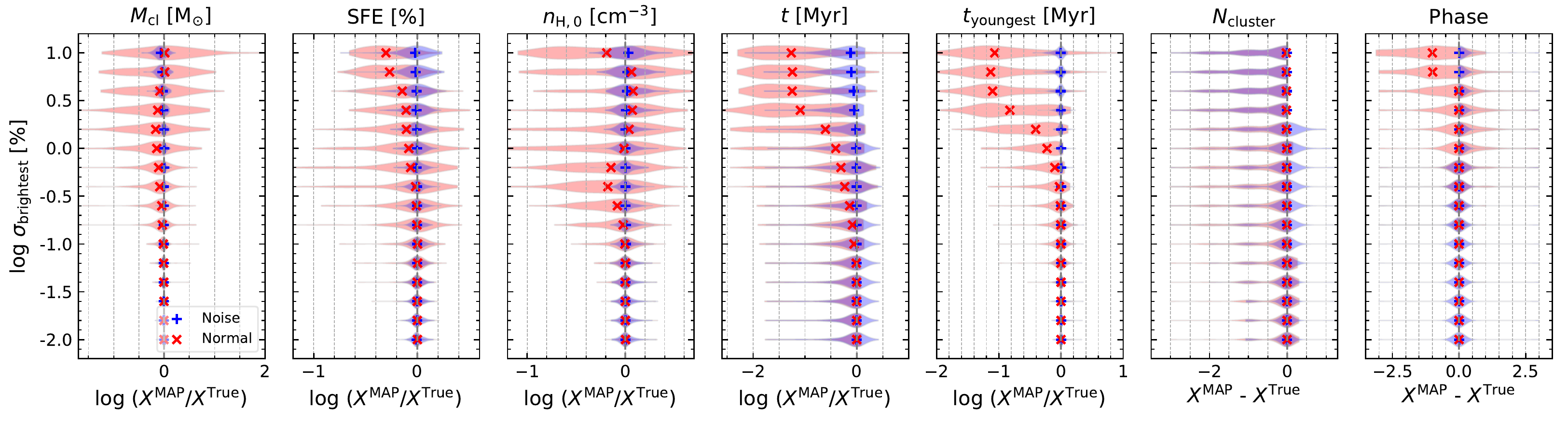}
	\includegraphics[width=2\columnwidth]{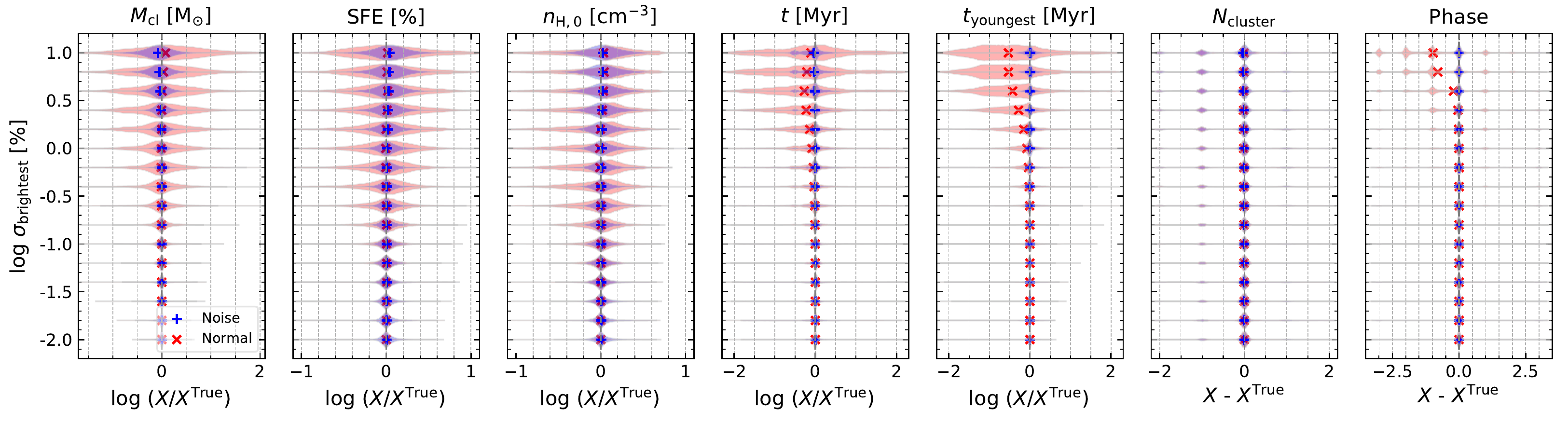}
    \caption{
    Histograms of two accuracy measures using 3200 posterior distributions (100 test models, 16 luminosity errors of the brightest emission line) obtained from the \noise\ (blue histograms) and the \normal\ (red histograms). We use the MAP estimates in the first row and use the entire posterior estimates in the second row. 
    The blue plus marks (\noise) and the red cross marks (\normal) indicate the median values of the histograms. We use the logarithmic deviation between the posterior estimates and the ground truth values of the test models except for \ncl\ and phase where we use the linear deviation.
     }\label{fig:violin}
\end{figure*}

Figure~\ref{fig:violin} shows the histograms of the two accuracy indices for the 16 different $\sigma_{\mathrm{b}}$ values for the \normal\ (red) and the \noise\ (blue). We use the MAP accuracy in the upper panels and the accuracy using the entire posterior estimates in the lower panels. 
In the upper panels, the MAP accuracy distribution of the \normal\ becomes wider and shifted with increasing error. The shift of the distribution is well described by the median value of the distribution denoted by the red x-shape mark. In particular, in the case of the first cluster age ($t$) and the youngest cluster age ($t_{\mathrm{youngest}}$), the median value is shifted by more than 1~dex toward negative value when the error is larger than 2.5\%\ (log $\sigma_{\mathrm{b}}$ > 0.4). This trend was already revealed in \citetalias{Kang+22}, where we found the \normal\ usually returned a very young age estimate with a single cluster when the error is large.

On the other hand, the prediction of the \noise\ is on average more accurate than the \normal\ for all seven parameters. The blue histogram is overall much narrower than the red histogram and the difference in width becomes more clear when the error is larger than 1\%. The accuracy distributions for \mcl\ and the youngest cluster age clearly show this trend. Contrary to the \normal, the median value of the \noise\, is always close to 0 without an offset even for the largest error of 10\%. The age of the first cluster is the only parameter that shows a shift of the median value for a large error, but the offset of around 0.2~dex is still small compared to the \normal\, where the median value is shifted by about 1.3~dex at $\sigma_{\mathrm{b}}$ of 10\%. Unlike for the age of the first cluster, the \noise\ shows an overwhelmingly accurate prediction for the youngest cluster age, even at an error of 10\%\ compared to the \normal.

The difference in the width between the blue and red histograms is noticeable as well when we evaluate the accuracy of the entire posterior estimates (lower panels in Figure~\ref{fig:violin}). The difference in the width is distinct when the error is large, especially in the case of \mcl, the oldest cluster age and the youngest cluster age. However, in the case of the density, the width of the \noise\ and the \normal\ appears fairly similar. The median value of the histogram mostly exhibits no offset for both \noise\ and \normal, except for the distributions of the youngest cluster age and phase of the \normal. 
Based on the two violin plots in Figure~\ref{fig:violin}, we discover that the predicted posterior distributions from both networks widen with increasing error, but the \noise\ results remain much narrower than the \normal. Furthermore, the \noise\ maintains the accurate MAP prediction at large error, whereas the \normal\ does not.

\begin{figure*}
	\includegraphics[width=2\columnwidth]{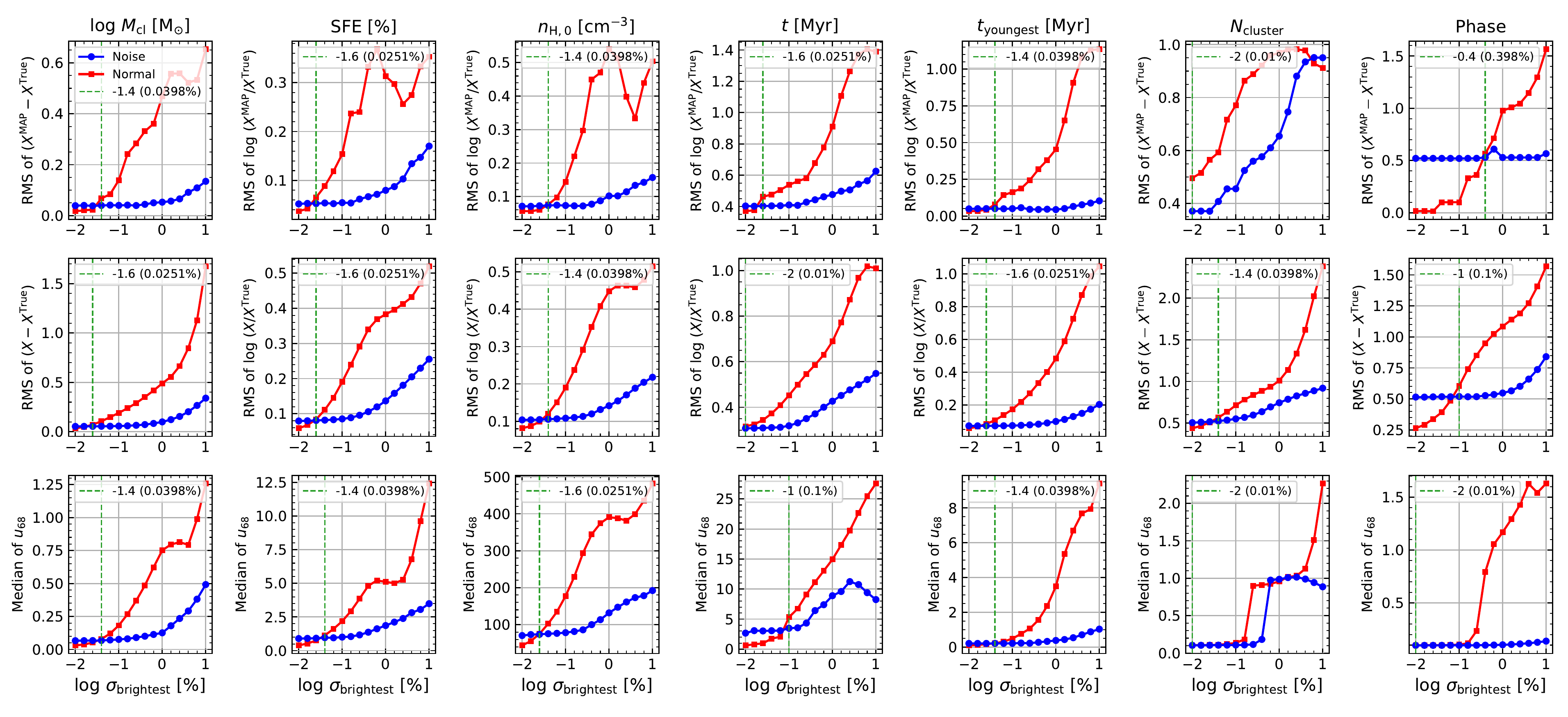}
    \caption{Accuracy and precision of the \noise\ (blue lines) and the \normal\ (red lines) as a function of the luminosity error of the brightest emission line ($\sigma_{\mathrm{brightest}}$) using 100 test models. We present the RMS of two accuracy measures: MAP accuracy in the first row and accuracy using all posterior estimates in the second row. The last row shows the median of the uncertainty at a 68\%\ confidence interval ($u_{68}$) which represents the precision of our network.
    Vertical green dashed lines show the error values at the turning points where the \noise\ begins to perform better than the \normal.
     } \label{fig:rmse}
\end{figure*}

Figure~\ref{fig:violin} reveals that the \noise\ performs better than the \normal\ overall, especially for large errors, and resolves the offset issues of the \normal. In Figure~\ref{fig:rmse}, we demonstrate how the accuracy and precision indices on average change with increasing error in a more quantitative way. The first two rows show the root mean square (RMS) value of the two accuracy indices, respectively, and the third row presents the median value of the precision index ($u_{68}$).

As already indicated by Figure~\ref{fig:violin}, Figure~\ref{fig:rmse} reveals that the \normal\ performs better than the \noise\ for small errors up until a critical error value is reached, where the \noise\ then surpasses the \normal. In each panel, we present the turning point, from which the \noise\ predicts better than the \normal, with the green vertical dashed line and in the legend. The turning point falls mostly around a value of 0.025\%\ -- 0.04\%\ (log $\sigma_{\mathrm{b}}$: -1.6 -- -1.4). In the case of phase accuracy, the turning point is larger than the other parameters. However, the RMS value of the \noise\ at the error smaller than the turning point is around 0.5 which is still acceptable.

The larger the error after the turning point, the larger the performance gap between the two networks. This is because the performance of the \normal\ deteriorates significantly as the error increases, whereas the performance change of the \noise\ is small. In the second row, the curves for \mcl\ and the youngest cluster age show the clear gap between the two networks where the accuracy differences between the two networks at the error of 10\%\ are 1.4 and 0.8~dex, respectively.

We measure the error when the performance of the \normal\ is comparable to the performance of the \noise\ at $\sigma_{\mathrm{b}}$ of 10\%. In most cases, the performance of the \noise\ at a 10\%\ error is similar to the performance of the \normal\ at an error of 0.4\%. In the previous paper, we demonstrated that the \normal\ provides reliable prediction when the error of the brightest line is smaller than 0.1$\sim$1\%. This result shows that the \noise\ works reliably even with large errors of around 10\%.

Figures~\ref{fig:violin} and \ref{fig:rmse} demonstrate that the \noise\ works significantly better than the \normal\ after the turning point located at a small error of around 0.04\%. It is also noteworthy that the offset issues and the difficulty of predicting ages at large errors in the \normal\ improves a lot in the \noise.

\subsection{Individual posterior distribution} 
\label{subsection:individual}

\begin{figure*}
	\includegraphics[width=1.6\columnwidth]{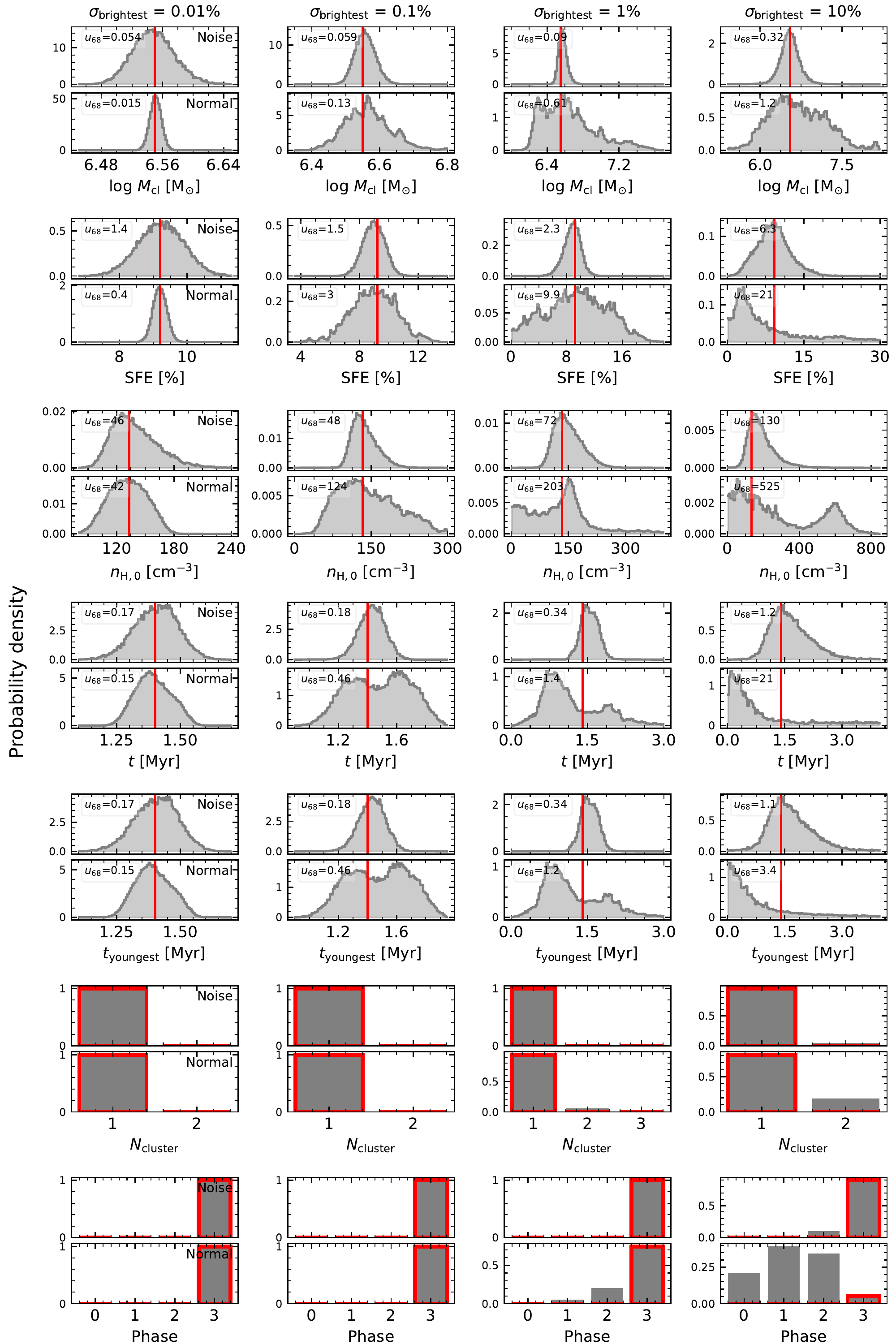}
    \caption{Posterior probability distributions (grey histograms) of the seven physical parameters of the first example model estimated by the \noise\ and the \normal\ for four different errors of the brightest emission line values (0.01, 0.1, 1 and 10\%). Each column corresponds to the result for each error. Please note that the range of the x-axis is different in all columns. Each panel is divided into two: the posterior distribution from the \noise\ (upper sub-panel) and the posterior distribution from the \normal\ (lower sub-panel). The red vertical lines indicate the true value of the example model. In the upper left corner, we present the $u_{68}$ value, the width of the distribution.
     } \label{fig:post_74}
\end{figure*}

\begin{figure*}
	\includegraphics[width=1.6\columnwidth]{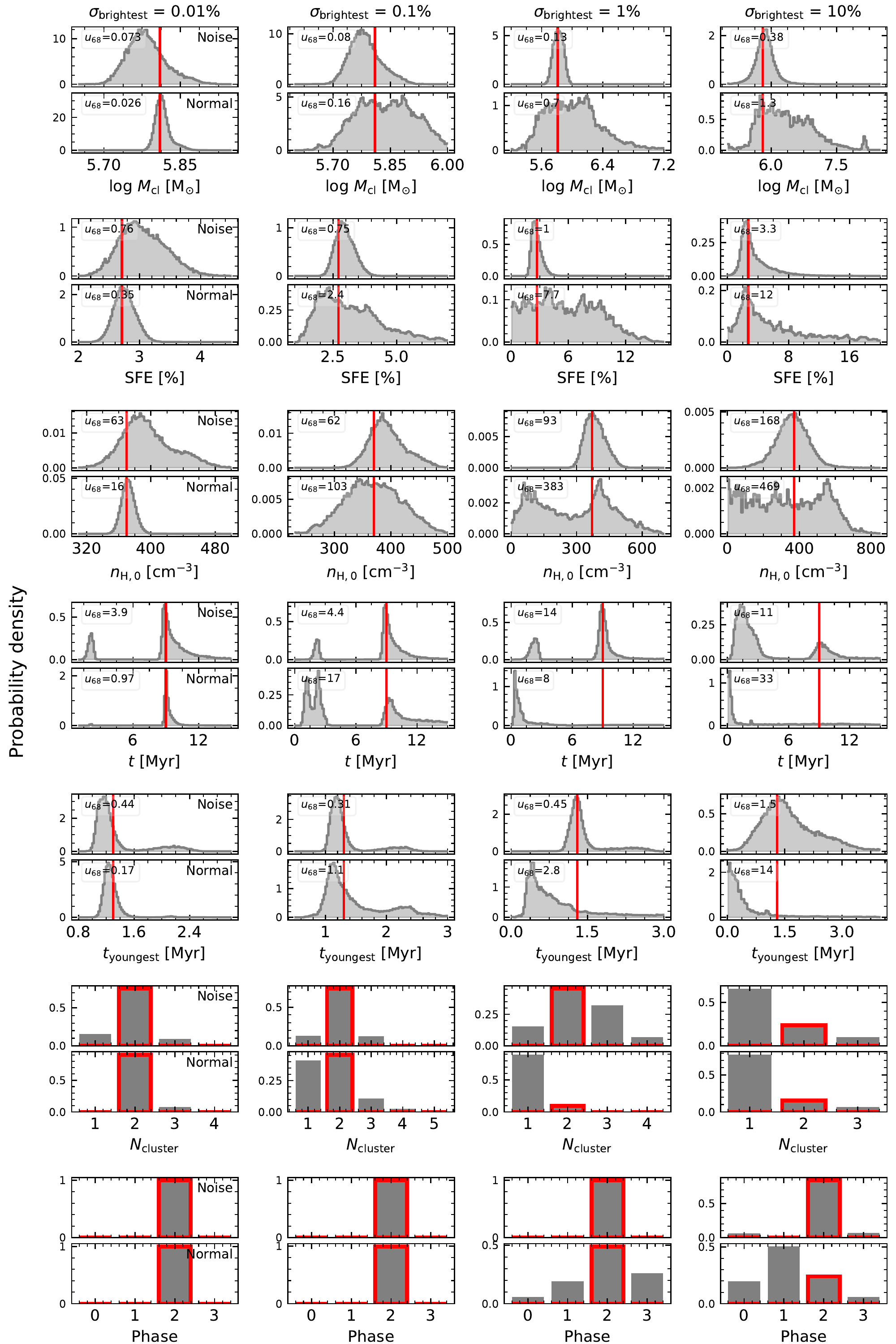}
    \caption{The posterior distributions of the second example model. Colour codes and lines are the same as in Figure~\ref{fig:post_74}.
     } \label{fig:post_33}
\end{figure*}

In this section, we compare the one-dimensional posterior distributions of the \normal\ and the \noise\ on two test models, which have a typical shape of posterior distributions.

The first example is a model at the age of 1.4~Myr in Phase 3, meaning that the expanding shell swept away all of the gas in the natal cloud. As there is only one cluster inside, the age ($t$) and the age of the youngest cluster ($t_{\mathrm{youngest}}$) are the same.
In Figure~\ref{fig:post_74}, we select four $\sigma_{\mathrm{b}}$ values (0.01, 0.1, 1 and 10\%) and present the corresponding predicted posterior distributions in each column. Each panel shows the posterior distribution of the \noise\ in the upper sub-panel and the posterior distribution of the \normal\ in the lower one.

The first column clearly shows that when the error is small (around 0.01\%) the posterior distribution of the \normal\ is narrower than the \noise. The difference is especially apparent in the case of \mcl\ and the star formation efficiency. Both networks predict accurately but the \normal\ performs slightly better for some parameters like \mcl\ and SFE. 
However, the situation is already reversed at an error of 0.1\%. The posteriors predicted by the \normal\ are wider than the \noise\ for all five continuous parameters. The posterior distributions of the \normal\ become less smooth but the corresponding MAP estimates are still accurate except for the two age parameters now showing bimodal distributions. On the other hand, the posterior of the \noise\ shows a unimodal distribution similar to the case at 0.01\%\ error and maintains accuracy although the width of the distribution does slightly increase.

As the error increases to 1 and 10\%, the posterior distributions predicted by the \normal\ for all parameters except \ncl\ and phase change significantly. The posterior distributions are spiky and irregular and for the ages, they now exhibit an offset towards lower values similar to the one in the violin plots (Figure~\ref{fig:violin}). While the network still predicts \ncl\ well, the phase estimation becomes degenerate, especially when the error is 10\%. In contrast, the posterior distribution of the \noise\ appears almost completely unaffected even when the error increases to 10\%. Only the width gradually increases but the MAP estimate remains accurate. At the largest error, the phase prediction does become degenerate but the MAP estimate is still accurate with a 90\%\ probability.

For the second example, we pick a model in a different evolutionary stage. The second model is in Phase 2 and has two clusters due to the stellar feedback of the first generation cluster being weaker than gravity, resulting in a recollapse. The age of the older generation cluster is 9~Myr and that of the younger one is 1.3~Myr.
At the lowest error (0.01\%, the first column in Figure~\ref{fig:post_33}), the posterior distributions are similar to those of the first example. The \noise\ shows slightly wider and less accurate distributions than the \normal. The difference to the first example is that both networks have a weak degeneracy in the \ncl\ estimation despite the small error. This behaviour is expected, because in \citetalias{Kang+22} we showed that \ncl\ is the most degenerate parameter and induces degeneracies in the other parameters as well. Moreover, degeneracy is more likely to occur when the target object has more than one cluster. As shown in the age posterior distribution of the \noise, the degeneracy in the \ncl\ leads to a degenerate prediction in the age of the first cluster.

If the error increases to 0.1\%, the posterior estimates of the \normal\ show wide distributions and the fraction of degenerate predictions increases in \ncl\ and age, whereas the posterior distributions of the \noise\ are almost identical to the result for the 0.01\%\ error. From an error of 1\% onwards, most of the posterior estimates of the \normal\ show a wide and skewed distribution similar to the first example, losing accuracy in most parameters. In the previous paper, we demonstrated that the \normal\ tends to predict extremely young age with a single cluster regardless of the target objects if the error is large (around 10\%). This means that the prediction of the \normal\ at a large error is unreliable and cannot be explained as physically valid alternatives to the target system. Contrary to the \normal, the posterior distributions predicted by the \noise\ remain accurate and narrow despite the large error. Although the fraction of the degenerate estimates (i.e., \ncl\ of 1) increases in \ncl\ and the age if the error is large (10\%), the other five parameters still have very accurate predictions with a unimodal distribution.

In many test models including these two examples, we confirmed that the performance of the \noise\ is inferior to the \normal\ for errors between 0.01 and 0.1\%, which is consistent with the turning point of 0.04\%\ found in the statistical evaluation (Figure~\ref{fig:rmse}). Additionally, the accuracy and precision of both networks deteriorate with increasing error, but the degree of deterioration is significantly different and the posterior distributions change differently. The \noise\ does not return extremely wide or skewed posterior distributions like the \normal\ and maintains an accurate prediction even at the largest error although the fraction of degenerate predictions increases slightly.

\section{Discussion}
\label{section:discussion}

\subsection{Skewness and degeneracy}
\label{subsection:skewness and degeneracy}

\begin{figure*}
	\includegraphics[width=2\columnwidth]{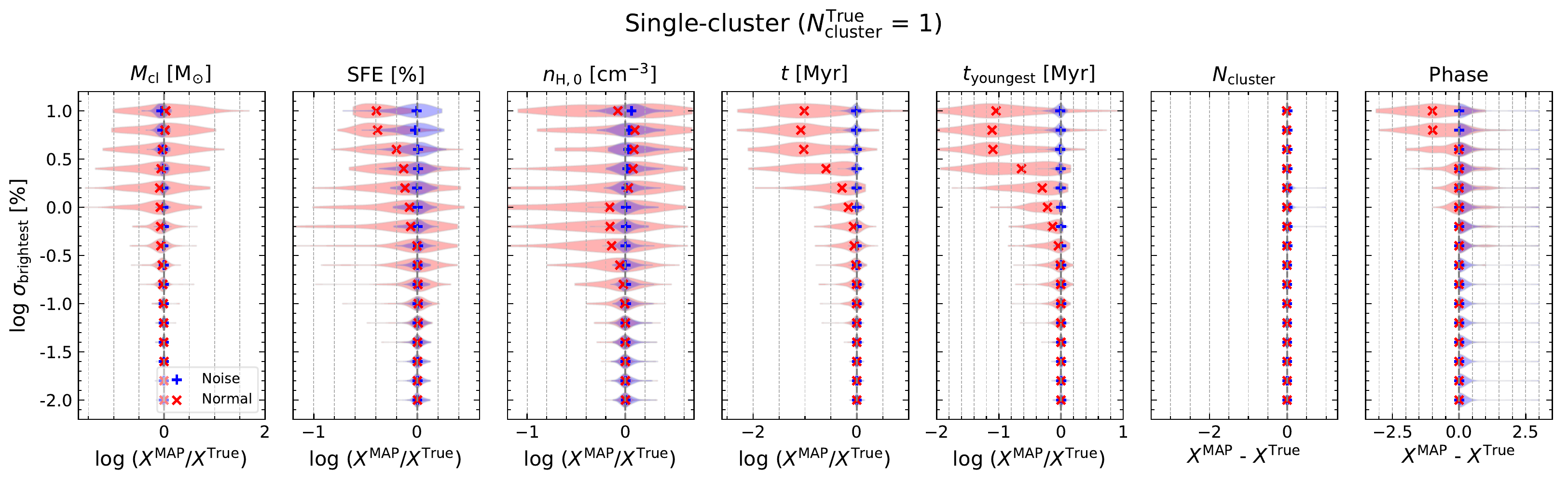}
	\includegraphics[width=2\columnwidth]{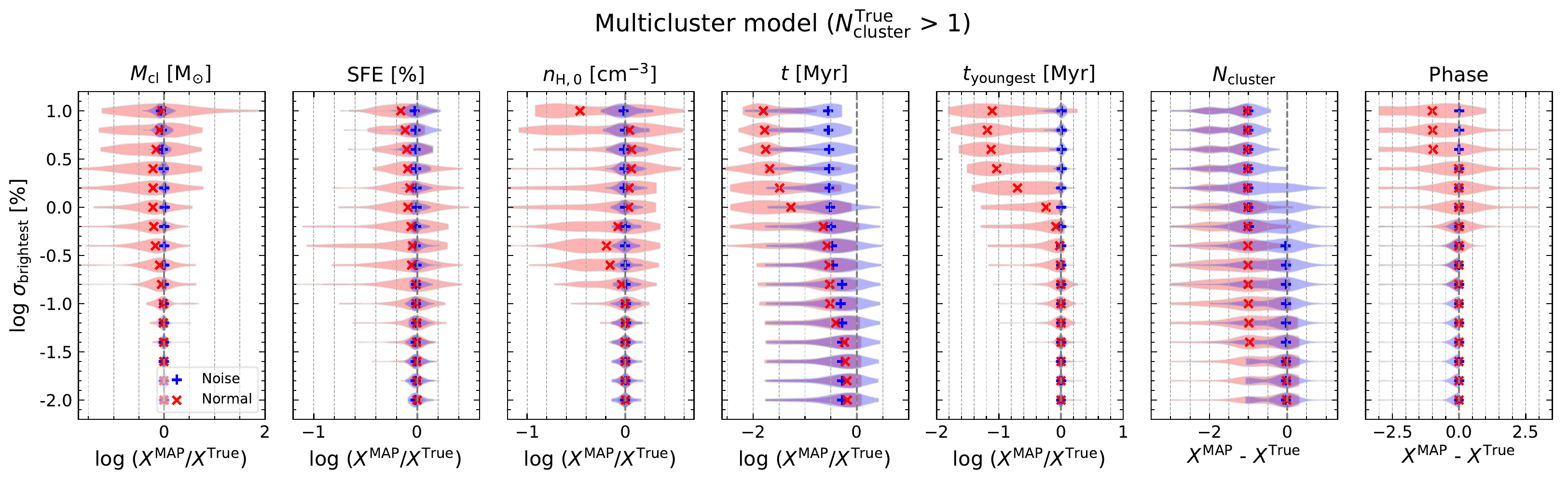}
    \caption{
    We divide the MAP accuracy violin plot in Figure~\ref{fig:violin} into two depending on the true \ncl\ value of test models: single cluster models (upper panels) and multicluster models (lower panels). 60 models have only one cluster and the other 40 models have more than one cluster. 
    Colour codes and symbols are the same as in Figure~\ref{fig:violin}.
     } \label{fig:violin_sm}
\end{figure*}  

If network prediction is degenerate, the posterior distribution shows multiple modes, one of which indicates the true value of the test model.
In \citetalias{Kang+22}, we demonstrated the modes other than the true mode in a degenerate posterior distribution are not wrong but are physically valid alternatives satisfying the same luminosity by re-simulating the posterior models via WARPFIELD-EMP and comparing the re-simulated luminosity with the true luminosity.
In this study, we do not re-simulate the posterior estimates. However, based on the results of \citetalias{Kang+22}, we regard that a multimodal posterior distribution reflects physically valid degeneracy remaining in the prediction, especially when the multimodality is due to the \ncl.

In \citetalias{Kang+22}, we showed that the most difficult parameters for our network to predict, even without considering errors, are the number of clusters (\ncl) and the age of the first generation cluster ($t$). We also found that more than 90\%\ of the degeneracy in the posterior distribution was caused by the degeneracy in \ncl. If the posterior of \ncl\ has a multimodal distribution, the posterior distribution of the age also has multiple modes in most of the cases, which lowers the (point estimate) accuracy. In particular, a degenerate \ncl\ prediction was more common if the target \hii\ region had multiple clusters, so the prediction was usually more accurate for single cluster models than multicluster models.

There are two main reasons why it is difficult to break the degeneracy in the \ncl\ prediction. One is the biased distribution of our training data. More than 70\%\ of our synthetic \hii\ region models have only one cluster and the age distribution is also biased towards a young age, so that our network usually learns better about young and single cluster models. The second reason is the selection of the emission lines. We use 12 optical emission lines that are tracers of ionized gas and that therefore are mostly influenced by the youngest cluster. The contribution of older clusters to the luminosity of these lines is often insignificant (see Figure 4 in \citealt{Pellegrini+20}) so it is hard for the network to identify how many old clusters are in the \hii\ region based only on these 12 lines.\footnote{Supplementing these diagnostics with additional observables sensitive to the older clusters (e.g., broad-band optical or near-IR colours) may allow many of these degeneracies to be broken, but is outside of the scope of our current study.}

As the networks in this study use the same database and the same emission lines, they essentially have the same difficulties. Therefore, we further examine the results presented in Section~\ref{section:result} by separating the sample into two groups, single cluster models and multicluster models, according to the true \ncl\ value of each model. In particular, we focus on the large error range ($\sigma_{\mathrm{b}} \geq 1$\%), where the \normal\ and the \noise\ begin to show a significant difference in average performance and individual predicted posterior distributions. According to the selection criteria in Section~\ref{subsubsection:sample selection}, 60 out of our 100 models have only one cluster, whereas the other 40 models have more than one cluster.
In Figure~\ref{fig:violin_sm}, we subdivide the violin plot of MAP accuracy from Figure~\ref{fig:violin} into single cluster models (upper panels) and multicluster models (lower panels), revealing a clear difference between the predictions of the \normal\ and the \noise\ at large error. 

Firstly, the prediction results of the \normal\ are similar between the two groups regardless of the true \ncl\ value except for the histograms of the \ncl\ estimation. When the error is larger than 1\%, the MAP estimates of the age of the oldest cluster, the age of the youngest cluster, and the phase are always notably offset towards smaller values in both groups. This feature is well revealed in the example posterior distributions (Figures~\ref{fig:post_74} and \ref{fig:post_33}) by the notable skewness. 
The difference in the \ncl\ histograms occurs, because the MAP estimates of the \normal\ for \ncl\ are always 1 when the error is large. Consequently, the histogram is close to 0 for single cluster models, because the true value is 1, but for the multicluster models, the MAP estimates are always smaller than the true value. In \citetalias{Kang+22}, we also confirmed that the \normal\ returns similar posterior distributions regardless of the target objects if the error is large, meaning that the \normal\ returns wrong estimates when subject to large errors. This is different to a true degeneracy in the prediction that shows the physically valid alternatives satisfying the same observation.

On the other hand, the \noise\ shows different performance for \ncl\ and the age of the first cluster depending on the true \ncl\ value. The \noise\ performs very well for all seven parameters in the case of single cluster models even at the large error of 10\%. The \noise\ predicts the age of the youngest and oldest cluster and the phase well unlike the \normal\, which shows systematic offsets. In the case of multicluster models, however, the \noise\ also exhibits an offset towards lower values in the oldest cluster age and \ncl, but it recovers the age of the youngest cluster and phase well. 
In the case of the \ncl\ prediction, the \noise\ also determines the MAP value as 1 in most cases regardless of the target model when the error is large ($\sigma_{\text{b}} > 1$\%). While the estimates of the oldest cluster age are also shifted towards lower values for the multicluster models even for the smaller error range, the magnitude of the shift is smaller than that of the \normal. When the error is small, both networks exhibit a shift of less than 0.5 dex, but when the error is large the \normal\ offset increases to about 2 dex, whereas the \noise\ shift remains at around 0.5 dex.

Based on Figure~\ref{fig:violin_sm} and the examples of the predicted posterior distributions (Figures~\ref{fig:post_74} and \ref{fig:post_33}), the likelihood of the \noise\ to return degenerate predictions increases with the error. However, unlike the \normal\ the \noise\ does not return wrong predictions. As previously mentioned, due to the intrinsic difficulty in predicting \ncl\ in our cINNs, the degeneracy in the \noise\ is mostly caused by degenerate \ncl\ predictions. As the prediction of the \noise\ is degenerate unlike the \normal, the estimation of the oldest cluster age by the \noise\ is not extremely offset towards younger models but shifted close to the true value of $t_{\text{youngest}}$. In addition, the \noise\ always returns the correct information about the youngest cluster and its evolutionary phase for both single cluster models and multicluster models.

While degenerate predictions are less accurate in terms of finding the unique true value, they are still meaningful in that they suggest other possible solutions that satisfy the same observation.
As most of the degeneracy lies within the \ncl\ prediction, we can filter out the correct posterior estimates if we can constrain the \ncl\ by another method or observation. We expect that by using more unbiased training data or including emission lines that are contributed by old clusters, we can improve the degeneracy in our networks.

\subsection{Pros and cons of \noise\ and \normal}
\label{subsection:pros and cons}

\renewcommand{\arraystretch}{1.25}
\begin{table*}
    \caption{ The average performance of the \noise\ and the \normal\ for 100 test models when $\sigma_{\text{b}}$ is 0.1\%. The first value in each entry shows the performance of the \noise\, whereas the second value is that of the \normal. The MAP accuracy denotes the deviation between the MAP estimates and the true value, while the accuracy in the second row refers to the deviation of all posterior estimates from the true value. We use a linear deviation for \ncl\ and phase and use a logarithmic deviation for the other five parameters.
    The last row indicates the average width of the posterior distribution. The listed values are the same as the data points in Figure~\ref{fig:rmse} at $\sigma_{\text{b}}$ of 0.1\%.
    \label{table:performance_0p1percent}
    }
    \begin{tabular}{ l  c  c  c  c  c  c   c }
        \hline
        \hline
        Performance index at $\sigma_{\text{b}} = 0.1$\%\ & log \mcl\ & SFE & \hdenini\ & $t$ & $t_{\text{youngest}}$ & \ncl\ & Phase \\
        (\noise\ / \normal)    \\ 
        \hline
        RMS of MAP accuracy & 0.04 / 0.14 & 0.055 / 0.15 & 0.073 / 0.14 & 0.41 / 0.54 & 0.05 / 0.16 & 0.46 / 0.77 & 0.52 / 0.1  \\ 
         RMS of accuracy & 0.058 / 0.19 & 0.085 / 0.19 & 0.11 / 0.19 & 0.32 / 0.45 & 0.074 / 0.17 & 0.55 / 0.71 & 0.52 / 0.6  \\ 
        Median of $u_{68}$ & 0.076 / 0.18 & 1 / 2.2 & 78 / 180 & 3.5 / 5.3 & 0.22 / 0.49 & 0.11 / 0.14 & 0.098 / 0.1  \\ 
        \hline
        \hline
    \end{tabular} 
\end{table*}

In Section~\ref{section:result}, we directly compared the performance of the \noise\ and the \normal. In this section, we compare the methodological differences between the two networks and discuss which network is more practical when applied to real observational data based on the results we presented.
The two networks differ in the method of training the network and sampling the posterior estimates. The advantages and disadvantages of the two networks that arise from the methodological aspect can be summarised as follows.

In the case of the \normal, we train the network using pure training data. To consider the observational error in the posterior distribution, $p(\mathbf{x}|\mathbf{y}, \boldsymbol{\sigma})$, we have to introduce the additional Monte Carlo-based marginalisation method (Eq.~\ref{eq:marginalisation}). We generate a sufficient number of perturbed luminosity sets by adding random Gaussian noise, assuming that the luminosity errors follow a Gaussian profile. As this method uses a trained network, we are free to choose the error profile to marginalise in this method. It is relatively easy to apply different profiles such as an asymmetric error profile or consider physical relationships between emission lines if needed depending on the object of interest. However, the disadvantage of this method is that it requires a lot of posterior estimates to fully sample the posterior distribution. Generating perturbed luminosity sets ($N_{\text{p}}$) and sampling the latent variables for each luminosity set ($N_{\text{z}}$), we generate $N_{\text{p}} \times N_{\text{z}}$ posterior estimates to make one posterior distribution. As it is important to use a sufficiently large $N_{\text{p}}$, we sample about 100,000 posterior estimates per one test model in this paper. This increases the overall computation time for posterior prediction and data analysis when using the \normal.

On the other hand, the methodological advantages and disadvantages of the \noise\ are opposite to the case of the \normal. We can obtain a full posterior distribution with only a small number of posterior estimates because the \noise\ does not require a marginalisation process (see Table~\ref{table:n_sample}). In other words, the \noise\ is advantageous in terms of the computation time for sampling the posterior estimates and processing the data. The disadvantage of the \noise\ is that we have to set the range of the luminosity error and the error profile in advance when training the network. Consequently, if we want to apply error values that are smaller or larger than the trained ranges or use a different error profile, we have to train a new network. In this paper, for simplicity, we assume that the errors of all 12 emission lines are independent to each other when using both \noise\ and \normal. If we want to consider any covariance between the lines, then we need to train a new \noise\ that applies the relations between the lines during the training.

Methodologically, \noise\ and \normal\ have advantages and disadvantages that are opposite to each other. However, based on the results in Section~\ref{section:result}, which network has better performance depends on the amount of the error. According to Figure~\ref{fig:rmse}, the \noise\ begins to perform better than the \normal\ on average when the error of the brightest emission line ($\sigma_{\text{b}}$) is larger than 0.04\%. At $\sigma_{\text{b}}$ of 0.1\%, the performance difference between the two networks is already noticeable and this performance gap widens as the error increases. This means that, if we apply our cINNs to real observations, the \noise\ is a more suitable method in terms of performance when the observation error corresponding to the $\sigma_{\text{b}}$ in our experiment is not less than 0.04\%.

To determine which network is more practical considering the error of real observations, we refer to the \hii\ region catalogue of \cite{Santoro+22} based on the PHANGS-MUSE survey~\citep{Emsellem+22}. In this catalogue, \cite{Santoro+22} lists observations for about 23000 \hii\ regions in 19 nearby spiral galaxies including the fluxes and corresponding errors for a set of strong lines observable within the wavelength range 4850--7000\AA. In our experiment in Section~\ref{section:result}, we use the error of the brightest line, i.e., the smallest error among the 12 emission lines by definition following Eq.~\ref{eq:luminosity noise model}. We confirm that, in the \hii\ region catalogue, both the brightest line and the line with the smallest flux error in per cent unit is always the \ha\ emission line. This is because the nebulae with emission lines brighter than the \ha\ were usually located outside of the \hii\ region area in the BPT diagram and excluded in the catalogue. The error of the \ha\ line in the catalogue is $0.302 \pm 0.294 \% $ on average and the minimum and maximum value are 0.011 and 3.08\%, respectively. 98\%\ of the \hii\ regions have an \ha\ error larger than 0.04\%\ and 87\%\ of them have an \ha\ error larger than 0.1\%. In other words, for most of the \hii\ regions in this survey, we expect the \noise\ to perform better than the \normal.

Based on the typical error of the real observations, we list the average performance (two accuracy indices and the precision index) at a $\sigma_{\text{b}}$ of 0.1\%\, presented in Figure~\ref{fig:rmse}, in Table~\ref{table:performance_0p1percent} to quantitatively examine the performance gap between the two networks. The first entry denotes the performance of the \noise\, while the second lists that of the \normal. This demonstrates that the \noise\ performs better than the \normal\ in all indices for all seven parameters. Excluding the phase and \ncl, the accuracy gap between the two networks is about 0.1 dex in each parameter. This suggests that, for the 85 per cent of \hii\ regions in \cite{Santoro+22}'s catalog, the \noise\ will at least give better results within this performance gap. As the average \ha\ error of the catalog is 0.3\%, the average performance gap will be larger than Table~\ref{table:performance_0p1percent} based on Figure~\ref{fig:rmse}.

In conclusion, \normal\ and \noise\ have opposite advantages and disadvantages in terms of methodology, but considering the range of the luminosity errors in practice, we expect that the \noise\ is a better choice than the \normal. We only gave one survey example but as we have the fiducial error to determine which method performs better, users can choose their best option considering their observational data. If it is reasonable to use identical error profiles for different \hii\ regions of interest and the error of the brightest line (usually the \ha) is sufficiently larger than 0.04 per cent, we propose that \noise\ would be a more robust method.

\subsection{Error larger than the training range}
\label{subsection:clipping}
The \noise\ learns about the observation error within the fixed range during the training. As mentioned in Section~\ref{subsection:noise_training}, we trained the \noises\ in this paper on errors ranging from 0.001 to 31.6~\%\ and applied the same error range for all 12 emission lines. However, in our experiment in Section~\ref{section:result}, there are many cases where the luminosity error exceeds the maximum training error (i.e., $\sigma_{\text{max}}$) of 31.6\%, especially when the $\sigma_{\text{b}}$ is larger than 1\%. When $\sigma_{\text{b}}$ is 1\%, 28 of 100 test models have an emission line whose luminosity error is larger than  $\sigma_{\text{max}}$. When $\sigma_{\text{b}}$ is 10\%, at least one emission line has an error larger than $\sigma_{\text{max}}$ in all test models.
While the \noise\ can somehow handle errors larger than the maximum it has learned, we need to examine if the \noise\ processes the error properly.

We hypothesize that the \noise\ either handles the large errors that it has never learned properly, or simply self-clips large errors to the training limit ($\sigma_{\text{max}}$). To investigate this hypothesis, we re-sample the posterior estimates for the 100 test models for 16 different $\sigma_{\text{b}}$ values as we did in Section~\ref{section:result}, but clip the luminosity error to $\sigma_{\text{max}}$ if it is larger than the maximum error. We analyze the newly obtained posterior distributions in the same manner as in Section~\ref{subsubsection:evaluation} and compare the result with the original outcome without error clipping from Section~\ref{section:result}.

Figure~\ref{fig:soft_clip} shows the difference between the unclipped original result (i.e., blue curve in Figure~\ref{fig:rmse}) and the new result with error-clipping. Differences begin to appear around a $\sigma_{\text{b}}$ value of 1\%, but the deviation is very small. Even when $\sigma_{\text{b}}$ is 10\%, the deviation of the two results is less than 0.02~dex in accuracy indices. The difference in median $u_{68}$ value is also very small considering the physical unit of the parameter. The difference between the original result without clipping and the result after clipping large errors to the maximum value is almost negligible.

\begin{figure}
	\includegraphics[width=1\columnwidth]{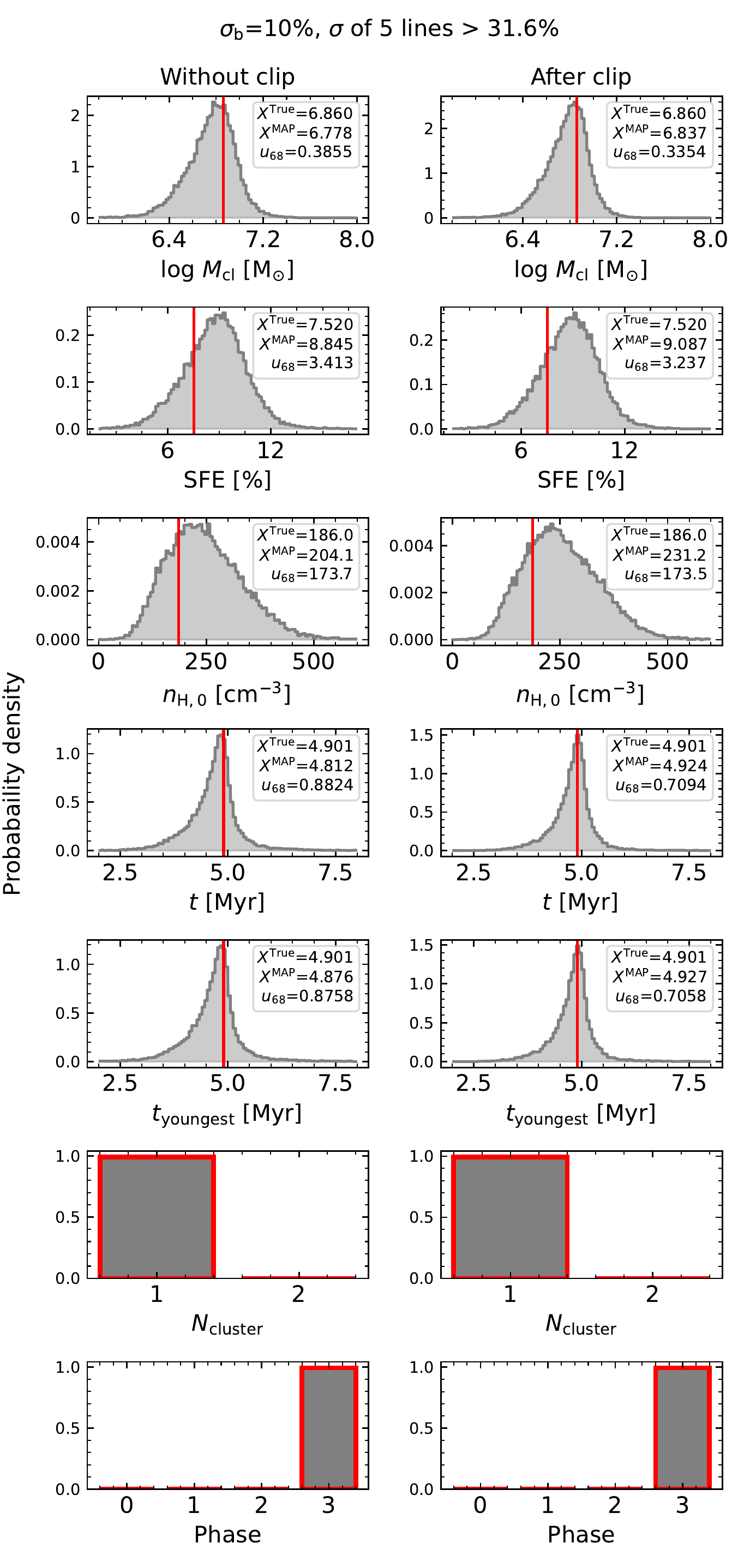}
    \caption{Predicted posterior probability distributions (grey histograms) for the seven physical parameters for one test model estimated by the \noise\ at $\sigma_{\text{b}}$ of 10\%. For this model, the luminosity errors of 5 emission lines are larger than the maximum error of the training range (31.6\%). The left panels show the posterior distribution using the luminosity errors without clipping, whereas the right panels present the posterior distribution after clipping the large errors to the maximum error. The red vertical lines indicate the true value of the model. In the upper right corner the true value, the MAP estimate, and the $u_{68}$ value of the posterior distribution are listed.
     } \label{fig:postclip_90}
\end{figure}

In addition to the average performance differences, we present one example to investigate if there is any noticeable difference in the posterior distributions for individual test models. In Figure~\ref{fig:postclip_90}, we present the predicted posterior distributions for one test model at a $\sigma_{\text{b}}$ of 10\%. In this case, 5 of 12 emission lines have a luminosity error larger than the maximum value (\oii\ 3726\AA, \oi\ 6300\AA, \sii\ 6716\AA, \sii\ 6731\AA\ and \sii\ blend 6720\AA). For the posterior distributions in the left panels, we use the large errors without clipping and for the right panels, we clip the 5 large error values to the maximum error of 31.6\%.
As shown in Figure~\ref{fig:postclip_90}, the two distributions are almost the same. While the left distributions without clipping appear a bit wider based on the $u_{68}$ values, the magnitude of the deviation is negligible.

Figures~\ref{fig:soft_clip} and \ref{fig:postclip_90} demonstrate that the \noise\ handles errors larger than its training range by self-clipping the large errors close to the upper limit of the training range. Accordingly, the direct comparison between the \noise\ and the \normal\ may not be fair because the \noise\ treats large errors as a smaller value by itself, while the \normal\ does not. Therefore, we also re-sample the posterior estimates using the \normal\ after clipping the large errors to the maximum error value and re-compare the new result of the \normal\ with the result of the \noise. We confirm that there is no change in the overall results presented in Section~\ref{section:result}. This is because the new result from the \normal\ as well is almost the same as the original result of the \normal\ without clipping. In Figure~\ref{fig:normal_clip}, we compare two results from the \normal: unlcipped original result (red curve in Figure~\ref{fig:rmse}) and the new result after clipping the large errors. Similar to the case of the \noise\ shown in Figure~\ref{fig:soft_clip}, the recognizable difference begins to appear around a $\sigma_{\text{b}}$ of 1\%. The differences between unclipped and clipped results are less than 0.03 dex in most cases except for the \mcl\ where the maximum difference is around 0.07 dex, which is still small enough. This implies that there is no significant difference in the posterior distribution from the \normal\ at large error range (i.e., larger than few tens per cent).

We propose the following reasons to explain the small differences in the results of the \normal. Firstly, at large errors, the \normal\ always returns a wide and skewed posterior distribution as shown in Figures~\ref{fig:post_74} and \ref{fig:post_33}. As mentioned in Section~\ref{subsection:skewness and degeneracy}, the \normal\ tends to predict similar posterior distributions at the largest error, regardless of the characteristics of the target model. The distribution is already as wide as the entire training range in Figure~\ref{fig:sample}, so even if we use a larger error, the posterior distribution does not widen any further or become more skewed. 
Secondly, in the marginalisation process for the \normal, we randomly perturb the luminosity by sampling from a Gaussian noise profile following Eq.~\ref{eq:perturb_lum}. However, we clip the perturbed luminosity to a minimum value of 1 to avoid unrealistic estimations. 

\begin{figure}
	\includegraphics[width=0.97\columnwidth]{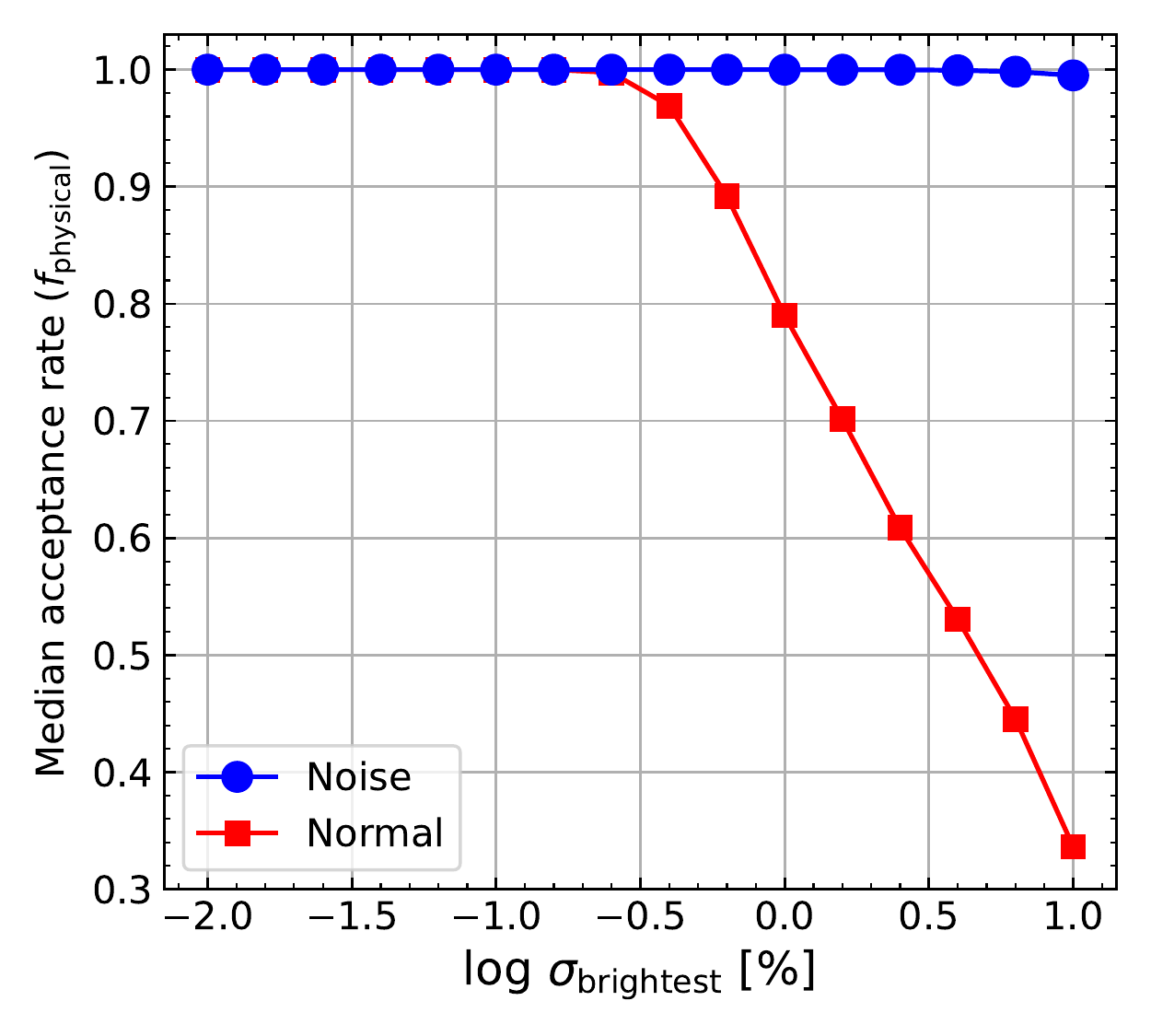}
    \caption{Median acceptance rate for 100 test models of the \noise\ and the \normal\ as a function of the luminosity error of the brightest emission line. The acceptance rate is the fraction of the posterior estimates that are not excluded as physically incorrect or extremely extrapolated values (e.g., negative age or star formation efficiency or age larger than 100~Myr) over the total number of the sampled posterior estimates.
     } \label{fig:acceptance}
\end{figure}

Thirdly, when using the posterior distribution for the analysis, we exclude all unrealistic posterior estimates such as negative densities or ages larger than 100~Myr. We define the acceptance rate ($f_{\mathrm{physical}}$) as the fraction of posterior estimates that are not regarded as unrealistic over the total number of the posterior samples. In Figure~\ref{fig:acceptance}, we compare the median acceptance rate for 100 test models of the two networks as a function of the luminosity error ($\sigma_{\mathrm{b}}$). Figure~\ref{fig:acceptance} clearly shows that the acceptance rate of the \normal\ significantly decreases with increasing error whereas the \noise\ maintains an acceptance rate close to 100\%. The acceptance rate of the \normal\ begins to decrease at an error of 0.25\% (log $\sigma_{\mathrm{b}}$ = -0.6) and decreases to 34\%\ for an error of 10\%. The acceptance rate of the \noise\ is always larger than 99.5\%. This result explains why the posterior distribution of the \normal\ does not change much at large error. Moreover, this demonstrates that the \noise\ provides physical estimates even when the error is large, although degeneracy remains in the posterior distribution, as discussed in Section~\ref{subsection:skewness and degeneracy}.


In summary, we find that the \noise\ clips large errors outside the training range by itself but this does not change the overall trend between the \noise\ and the \normal\ shown in Section~\ref{section:result}. Because the posterior distributions of the \normal\ do not change significantly at large error range. Considering the behaviour of the \noise\ and \normal\ at the large error range, it is better not to use our cINNs to analyze objects with too large luminosity errors above a few tens per cent.

\subsection{Outperformance and saturation of the \noise\ }
In Figure~\ref{fig:rmse}, we showed that the performance of the \noise\ and \normal\ change differently as a function of the observational error. For example, the RMS deviation of the \normal\ steadily increases from the minimum error of 0.01\%\ to the maximum error of 10\%. In contrast, the RMS deviation of the \noise\ gradually increases after 0.1\%, maintaining a small value in the small error range. Therefore, there is a turning point in which the performance of the \noise\ overtakes the \normal.

The turning point varies slightly depending on the parameter and performance index, but as mentioned in Section~\ref{subsection:statistical}, it mainly falls around a value of 0.025\%\ $\sim$ 0.04\%\ (log $\sigma_{\mathrm{b}}$: -1.6 $\sim$ -1.4). The performance of the two networks is similar at the turning point, but according to Table~\ref{table:performance_0p1percent} and Figure~\ref{fig:rmse}, the performance difference between the two networks is evident even at the error of 0.1\%. The larger the error, the larger the gap between the two networks.

We speculate that the reason why the \noise\ outperforms the \normal\ at large error is because of the differences in training methods. In the case of the \normal, the network learns from the same training data for each training epoch repeatedly during the training. On the other hand, the \noise\ learns about various luminosity and error values from the identical training models because the perturbations are randomly sampled at every epoch. This implicitly increases the number of training examples for the \noise\ even if the number of actual training models is the same as in the normal training. Furthermore, as the luminosity of the training model is perturbed by the randomly sampled noise during the noise training, the \noise\ learns more diverse cases of degeneracy and wider training distributions. For these reasons, the \noise\ performance deteriorates less than that of the \normal\ at a larger error.

On the other hand, the performance of the \noise\ does not change as much as the \normal\ when the observational error decreases. In Figure~\ref{fig:rmse}, the performance of the \noise\ appears to almost saturate at around 0.1\%\ and does not decrease much further with decreasing error. 
This also can be partly explained by the difference in the training methods. No matter how small the errors sampled during the noise training become, the \noise\ has never seen a version of the data as precise as the pure training data. Therefore, the predicted posterior distributions may have a basic degeneracy even at the minimum error, showing the saturation in Figure~\ref{fig:rmse}.
However, the \noise\ starts to saturate at around 0.1\%, which is a fairly large value considering the minimum training error of 0.001\%. The reason for the early saturation has not become clear within the scope of this study, but we will examine this trend in further experiments such as changing the profile of the error in the noise training in future works.

\section{Summary}
\label{section:summary}
In this paper, we introduce a new type of cINN, the \noise, that predicts physical parameters (\textbf{x}) of \hii\ regions from the emission-line luminosities (\textbf{y}), considering the uncertainty of the observations ($\boldsymbol{\sigma}$, emission-line luminosity error). 
In our recent paper (\citetalias{Kang+22}), we first introduced a cINN that predicts the seven physical parameters of the \hii\ region from the luminosity of 12 optical emission lines. The type of cINN used in \citetalias{Kang+22} is \normal\ which estimates posterior distributions of the physical parameters conditioned only on the luminosities ($p(\textbf{x}|\textbf{y})$), but we presented a method of considering luminosity errors in the \normal\ through a modification of the posterior sampling procedure (Section~\ref{subsection:sampling}) and showed the performance of the \normal\ as a function of the luminosity error. 
The \noise, newly introduced in this paper, always reflects the luminosity error in the prediction of the parameters by using both luminosities and corresponding errors as an input of the network ($p(\mathbf{x}|\mathbf{y}, \boldsymbol{\sigma})$). In this paper, we introduce the \noise\ and its training method (i.e., noise training) and compare the performance of the \noise\ with the \normal.

For the \noise\ to learn the influence of observational errors, the training method of the \noise\ is slightly different to the \normal. At each training epoch, we first randomly sample the errors of each emission-line luminosity from the prescribed probability distribution. In this paper, we sample the error in the logarithmic scale from the uniform distribution (Eq.~\ref{eq:sample_sigma}) with minimum and maximum errors of 0.001\%\ and 31.6\%, respectively. Next, we perturb the luminosities of the training data by adding the random Gaussian noises based on the sampled errors. Due to these two steps processed on the fly during the training, the \noise\ learns about the different training data every epoch whereas the \normal\ learns about the pure training data repeatedly.

We use synthetic \hii\ region models produced by WARPFIELD-EMP~\citep{Pellegrini+20} to train the cINNs because it is hard to collect lots of well-interpreted real data required for the training. WARPFIELD-EMP is a pipeline that calculates emission from isolated massive star-forming clouds using CLOUDY~\citep{Ferland+17} and POLARIS~\citep{Reissl+16, Reissl+19}, based on the one-dimensional semi-analytic stellar feedback code WARPFIELD~\citep{Rahner+17}.
The database of WARPFIELD-EMP models used in this paper is the same as the database introduced in \citetalias{Kang+22}. The database consists of 505,748 synthetic \hii\ region models evolved from 10,000 initial star-forming clouds. We use 90\%\ of the models to train the network and use the rest (test set) to evaluate the network performance.

In this paper, we present two cINNs,  one \noise\ and one \normal, that predict seven physical parameters from the information on 12 emission lines (see Table~\ref{table:param_obs}) trained on the WARPFIELD-EMP models. To compare the performance of the \noise\ and the \normal\ as a function of the luminosity error, we evaluate the accuracy and precision of two cINNs at 16 different levels of luminosity error using the error of the brightest emission line ($\sigma_{\mathrm{b}}$) as a representative error. Our main results of the comparison between the \noise\ and the \normal\ are the following:

\begin{enumerate}
    \item As the error increases, the performance of both networks gradually deteriorates, but the two networks show different trends as a function of the error. The \normal\ predicts more accurately and precisely than the \noise\ when the error is very small ($\sigma_{\mathrm{b}}$ $\sim$ 0.01\%). However, the performance of the \normal\ steadily degrades significantly compared to the \noise. 
    On the other hand, the \noise\ maintains good performance with increasing error at a small error range. The performance of the \noise\ slowly and gradually degrades at large errors.
    
    \item  From a certain point (i.e., turning point), the \noise\ outperforms the \normal\ and the larger the error, the larger the gap between the \noise\ and the \normal. The turning point occurs at an error of around 0.025--0.04\%.
    
    \item The \noise\ estimates parameters with sufficient accuracy and precision even when the luminosity error is large. The performance of the \noise\ when the error of the brightest line is 10\%\ is similar to that of the \normal\ when the error of the brightest line is only 0.4\%.

    \item The \noise\ predicts parameters much better than the \normal\ even when the error of the brightest line is 0.1\%.
    Considering the luminosity error of the observed \hii\ regions from the PHANGS-MUSE survey~\citep{Emsellem+22} as an example, the \ha\ luminosity error of 0.1\%\ is a small enough value because 89\%\ of the \hii\ regions had the \ha\ luminosity error larger than 0.1\%. This means that the \noise\ is a more appropriate tool when applied to real observations that have a similar level of uncertainty to the PHANGS-MUSE survey.
    
    \item As the error increases, degeneracy becomes common in the posterior distribution of the \noise.
    In the case of the \normal, the posterior estimates at a large error are not degenerate like the \noise\ but are unphysical. The fraction of the unphysical posterior estimates increases significantly as a function of the luminosity error in the case of the \normal, whereas the fraction for the \noise\ is almost zero even at the large error, meaning that the \noise\ always provides physically valid estimates although the degeneracy remains.

    \item The \noise\ learns about errors within a given training range (i.e., 0.001 -- 31.6 \%). If the \noise\ receives an error larger than the training range, the \noise\ self-clips the large errors close to the maximum of the training range and provides a posterior distribution using the clipped error.
    
\end{enumerate}

The \noise, newly presented in this paper, predicts parameters accurately and precisely even when the observation error is large.  
Our results of comparison between the \noise\ and the \normal\ will help select an appropriate network type according to the observed luminosity errors. Based on our results, we suggest utilizing the \noise\ if the error of the brightest emission line (e.g., the \ha\ line) is 0.1\% or more.

\section*{Acknowledgements}


We acknowledge funding from the European Research Council via the ERC Synergy Grant ``ECOGAL'' (project ID 855130), from the Deutsche Forschungsgemeinschaft (DFG) via the Collaborative Research Center ``The Milky Way System''  (SFB 881 -- funding ID 138713538 -- subprojects A1, B1, B2 and B8) and from the Heidelberg Cluster of Excellence (EXC 2181 - 390900948) ``STRUCTURES'', funded by the German Excellence Strategy. We thank the German Ministry for Economic Affairs and Climate Action for funding in the project ``MAINN'' (funding ID 50OO2206). We also thank for computing resources provided by the Ministry of Science, Research and the Arts (MWK) of the State of Baden-W\"{u}rttemberg through bwHPC and DFG through grant INST 35/1134-1 FUGG and for data storage at SDS@hd through grant INST 35/1314-1 FUGG.

\section*{Data Availability}

The code FrEIA used in this article as the framework of cINNs is available at \href{url}{https://github.com/VLL-HD/FrEIA}.
The database of WARPFIELD-EMP models and the code WARPFIELD-EMP used in this article will be shared on reasonable request to the corresponding author with the permission of Eric Pellegrini.

The data outcomes underlying this article will be shared on reasonable request to the corresponding author.




\bibliographystyle{mnras}




\appendix
\section{Supplemental materials}
\begin{figure*}
	\includegraphics[width=2\columnwidth]{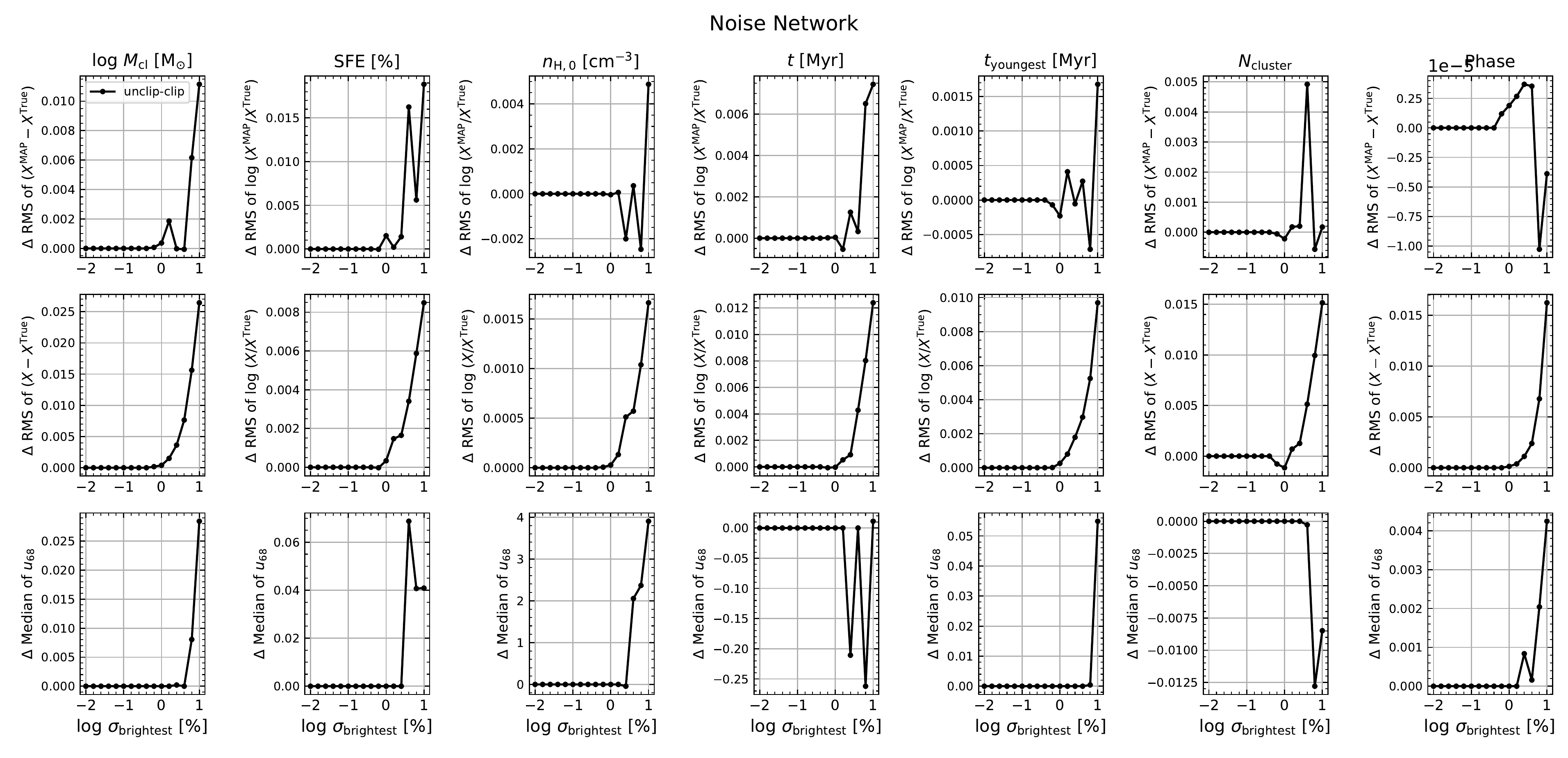}
    \caption{
     Average performance difference of the \noise\ between the original result (blue line in in Figure~\ref{fig:rmse}) and the results after error clipping. For errors larger than the training range of the \noise, we clip the error to the maximum value (31.6\%) and re-sample the posterior distributions for all test models. 
     }\label{fig:soft_clip}
\end{figure*}

\begin{figure*}
	\includegraphics[width=2\columnwidth]{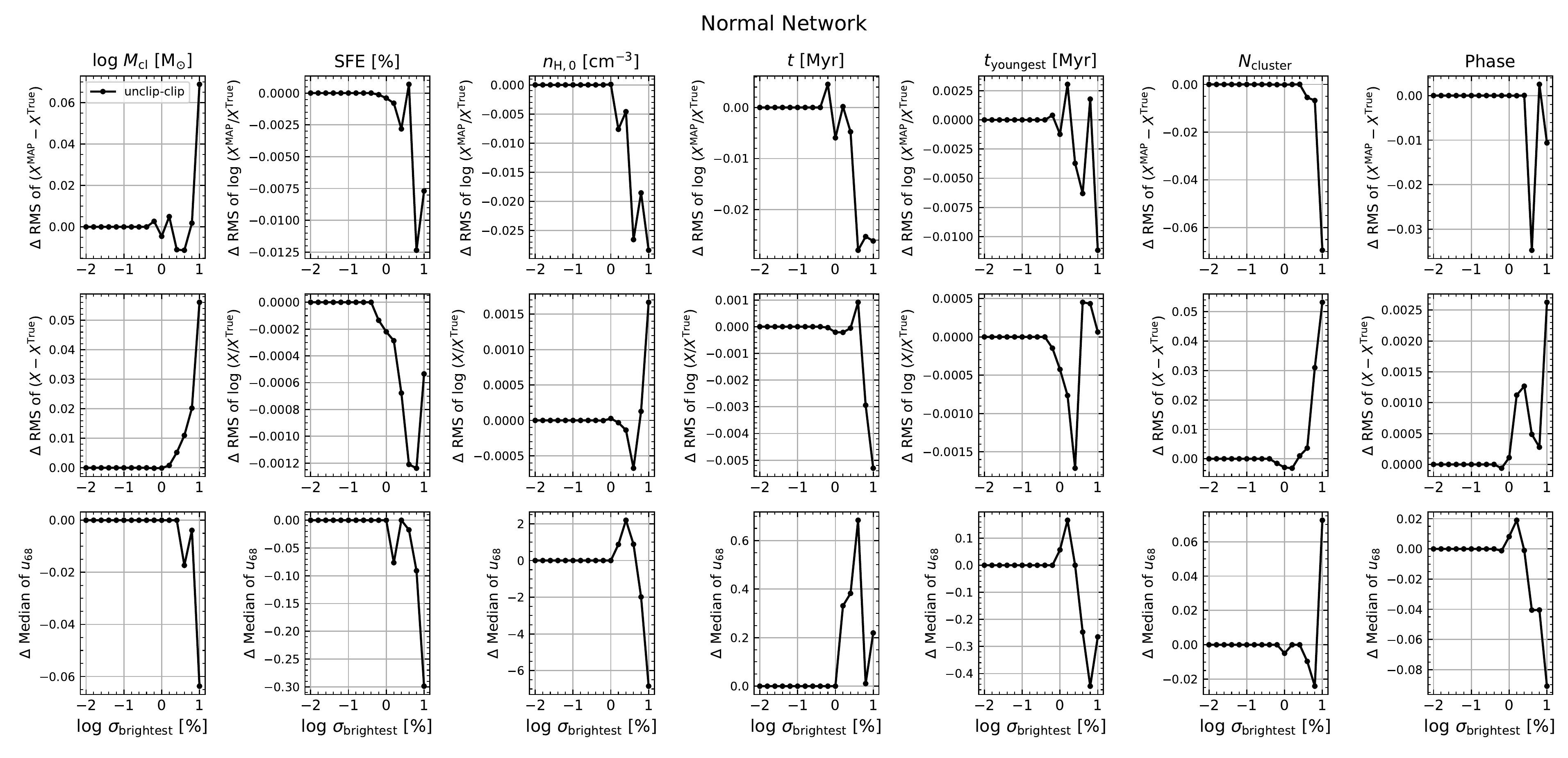}
    \caption{
    Similar to Figure~\ref{fig:soft_clip}, we present the average performance difference of the \normal\ between the original result (red line in Figure~\ref{fig:rmse}) and re-sampled posterior distributions after clipping the large errors. 
     }\label{fig:normal_clip}
\end{figure*}

In this appendix, we present supplementary figures for the error clipping tests for the  \noise\ and the \normal\ discussed in Section~\ref{subsection:clipping}. We re-sampled the posterior estimates for the 100 test models for 16 different $\sigma_{\mathrm{b}}$ levels after clipping the luminosity errors larger than the maximum training error ($\sigma_{\mathrm{max}}$) of 31.6\%. We present the average performance differences between the original unclipped result (Figure~\ref{fig:rmse}) and the clipped result of the \noise\ in Figure~\ref{fig:soft_clip} and of the \normal\ in Figure~\ref{fig:normal_clip}. As mentioned in Section~\ref{subsection:clipping}, the deviation between unclipped and clipped results is very small for both the \noise\ and the \normal.

\section{Deeper networks}
\label{section:deeper networks}
In this appendix, we provide the results of additional experiments that investigate the influence of network depth on prediction performance. Here the depth of the network refers to the number of affine coupling blocks ($N_\text{block}$) and the number of layers of the internal sub-network ($N_\text{layer}$). As the \noise\ processes more information than the \normal\ by definition, we investigate whether network performance improves by increasing the depth of the network. The networks used in the main paper are the best choices based on the results of this experiment.

In \citetalias{Kang+22}, we used eight affine coupling blocks and internal sub-networks with three layers to build the cINN ($N_\text{block}$=8, $N_\text{layer}$=3). Before the experiment, we confirmed that the \noise\ performed better than the \normal\ in general, similar to the results in Section~\ref{subsection:statistical}, even when we use the setup of \citetalias{Kang+22}. However, with the possibility of performance improvement in mind, we investigated the influence of $N_\text{block}$ and $N_\text{layer}$. 
To evaluate the network performance we used the same 100 test models and the same methodology as in Section~\ref{section:result}. For each network, we sample posterior distributions at 16 different $\sigma_{\mathrm{b}}$ for the 100 test models and measure two accuracy indices and the precision index for each posterior distribution. We compare the average performance of the network as a function of $\sigma_{\mathrm{b}}$ like the curves presented in Figure~\ref{fig:rmse}.

\begin{figure*}
	\includegraphics[width=2\columnwidth]{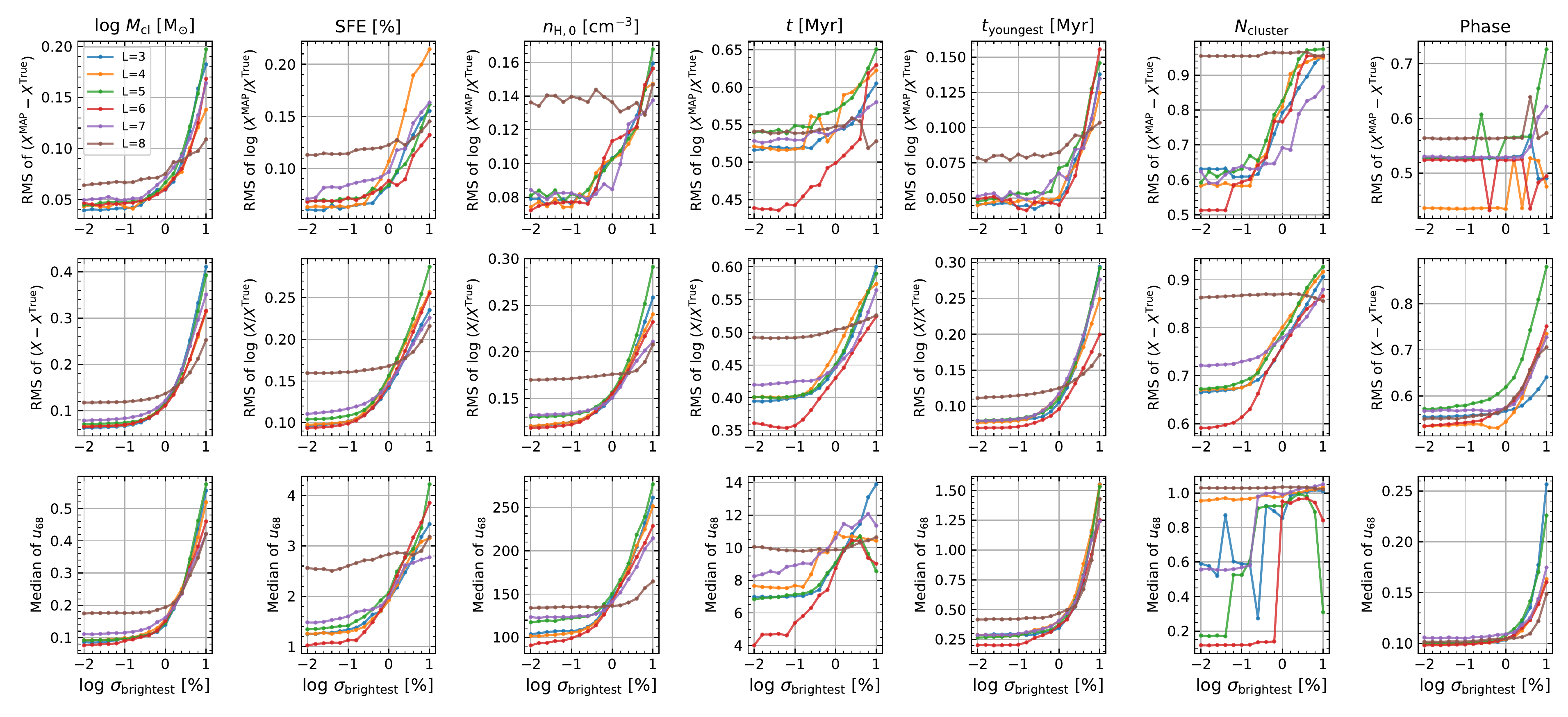}
    \caption{ 
    We compare the two accuracy indices (the first and the second row) and precision index (the third row) of six \noises\ for different numbers of layers in the internal sub-network. All six networks have 8 affine coupling blocks. We use the same sample of 100 test models and the same 16 different errors as used in Section~\ref{section:result}. 
     } \label{fig:layer_test}
\end{figure*}

In the first experiment, we train and compare 6 \noises\ by increasing the $N_\text{layer}$ from three to eight (Figure~\ref{fig:layer_test}). For these six networks, we fix $N_\text{block}$ at eight. Except for the worst network with eight layers (brown curve), the performance of the other networks is similar overall. It is not easy to identify the relation between the performance and the number of layers because the performance difference between networks slightly varies depending on the parameters and the range of the error. However, the performance gap between the network with six layers (red curve) and the rest is noticeable in the case of the oldest cluster age ($t$), especially when the error is small. For the other parameters, the red curve shows similar or slightly better results than the others. As we discovered in our first study (\citetalias{Kang+22}), the age of the oldest cluster is the most difficult parameter for our network to predict, so the improvement in $t$ prediction is meaningful. On this account, we choose the network with 6 layers as the best result of the first experiment. 

In Figure~\ref{fig:layer_test}, we present the results of 6 networks but we also tried to train the network with 9 layers. However, the latter experiment failed because the loss (Eq.~\ref{eq:loss}) did not decrease at all. In addition to the results for the networks with more than 6 layers in Figure~\ref{fig:layer_test} (purple and brown), this implies that increasing the number of layers does not always improve the prediction.

\begin{figure*}
	\includegraphics[width=2\columnwidth]{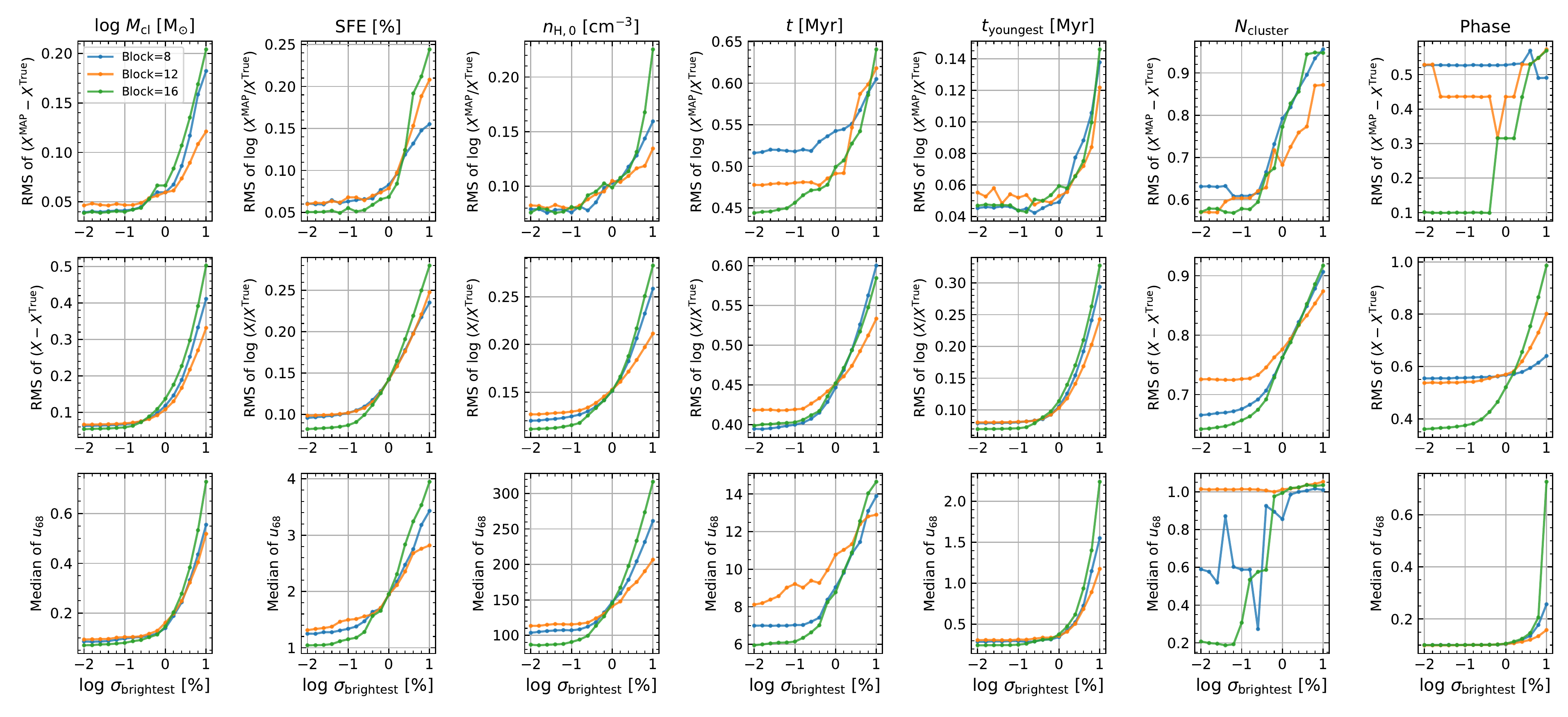}
    \caption{
    Comparison of the performance of 3 \noises\ with the different numbers of affine coupling blocks. All three networks use 3 layers in their internal sub-network.
     }\label{fig:block_test}
\end{figure*}

In the second experiment, we increase the number of affine coupling blocks ($N_\text{block}$) and fix the $N_\text{layer}$ to three. In Figure~\ref{fig:block_test}, we compare the performance of three networks with $N_\text{block}$ of 8, 12 and 16, respectively. All three networks perform similarly, but the network with 16 blocks (green curve) performs the best, especially in the case of the oldest cluster age ($t$). This network performs better than the other two at $\sigma_{\mathrm{b}}$ smaller than 1\%\ but performs poorly at large errors. As in the first experiment, we focus on improving the $t$ prediction and choose the network with 16 blocks as the best result.

\begin{figure*}
	\includegraphics[width=2\columnwidth]{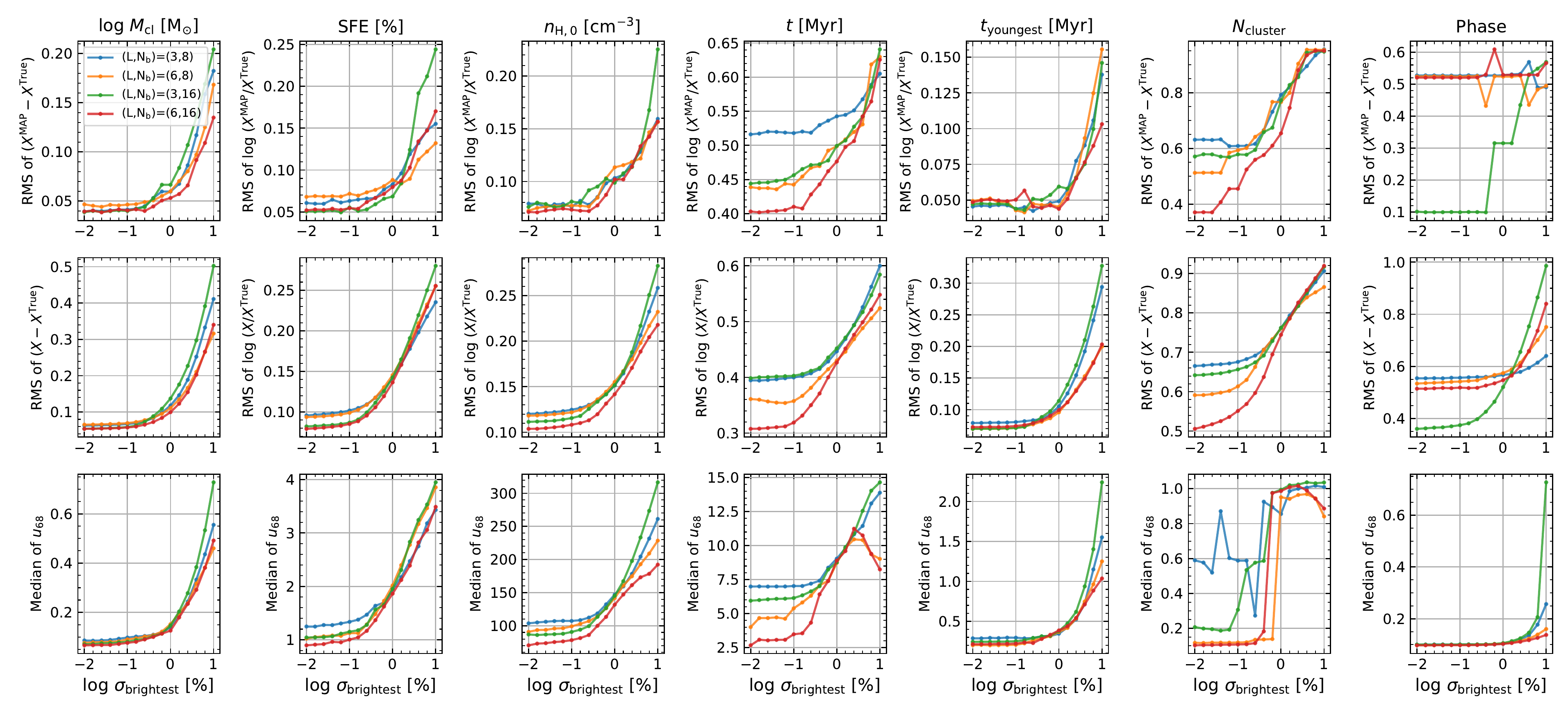}
    \caption{
    Comparison of the performance of four \noises\ with different combinations of the number of affine coupling blocks and the number of layers of the internal sub-network. The red line with 6 layers and 16 blocks is the same \noise\ used in the main paper (i.e., the blue line in Figure~\ref{fig:rmse}).
     }\label{fig:combi_test}
\end{figure*}

Lastly, we build the deepest cINN combining the best options from the previous two experiments ($N_\text{layer}$ = 6 and $N_\text{block}$ = 16) and compare the performance of the following four networks: the network with the \citetalias{Kang+22} setup ($N_\text{layer}$=3, $N_\text{block}$=8), the best network of the first experiment ($N_\text{layer}$=6, $N_\text{block}$=8), the best network of the second experiment ($N_\text{layer}$=3, $N_\text{block}$=16) and the deepest network ($N_\text{layer}$=6, $N_\text{block}$=16). In Figure~\ref{fig:combi_test}, the deepest network (red curve) performs the best in the case of \mcl, SFE, \hdenini\ and $t_\text{youngest}$, but the gap between the curves is not large. However, the deepest network shows a noticeable improvement in the oldest cluster age prediction and the accuracy of the \ncl\ prediction. In the case of the phase, the network with 3 layers and 16 blocks (green curve) performs the best, but considering the physical unit of the phase by definition, the performance of the other networks is still acceptable. Taking into account the overall performance, we choose the deepest network with 6 layers and 16 blocks as the best network and use this network for the main paper. So, the red curve in Figure~\ref{fig:combi_test} is the same as the blue curve in Figure~\ref{fig:rmse}.

Comparing the various \noises\ with different $N_\text{layer}$ and $N_\text{block}$, we found the following. Firstly, deepening the network can improve the prediction power, but it does not always guarantee improvement. Moreover, the performance does not change proportionally to the depth and training may fail if we deepen the network too much. Secondly, in the case of parameters that the cINN predicts relatively well (e.g., \mcl, $t_\text{youngest}$), there is no significant change depending on the network depth, but deepening the network improves the performance in the case of $t$ and \ncl\, which are relatively difficult for the cINN to accurately predict due to the degeneracy. In \citetalias{Kang+22}, we demonstrated that our cINN has difficulty resolving the degeneracy in \ncl\ prediction because of the biased parameter distribution of the training data and our selection of optical emission lines. This result implies that we can enhance the degenerate prediction of the \noise\ by properly increasing the network depth.

\begin{figure*}
	\includegraphics[width=2\columnwidth]{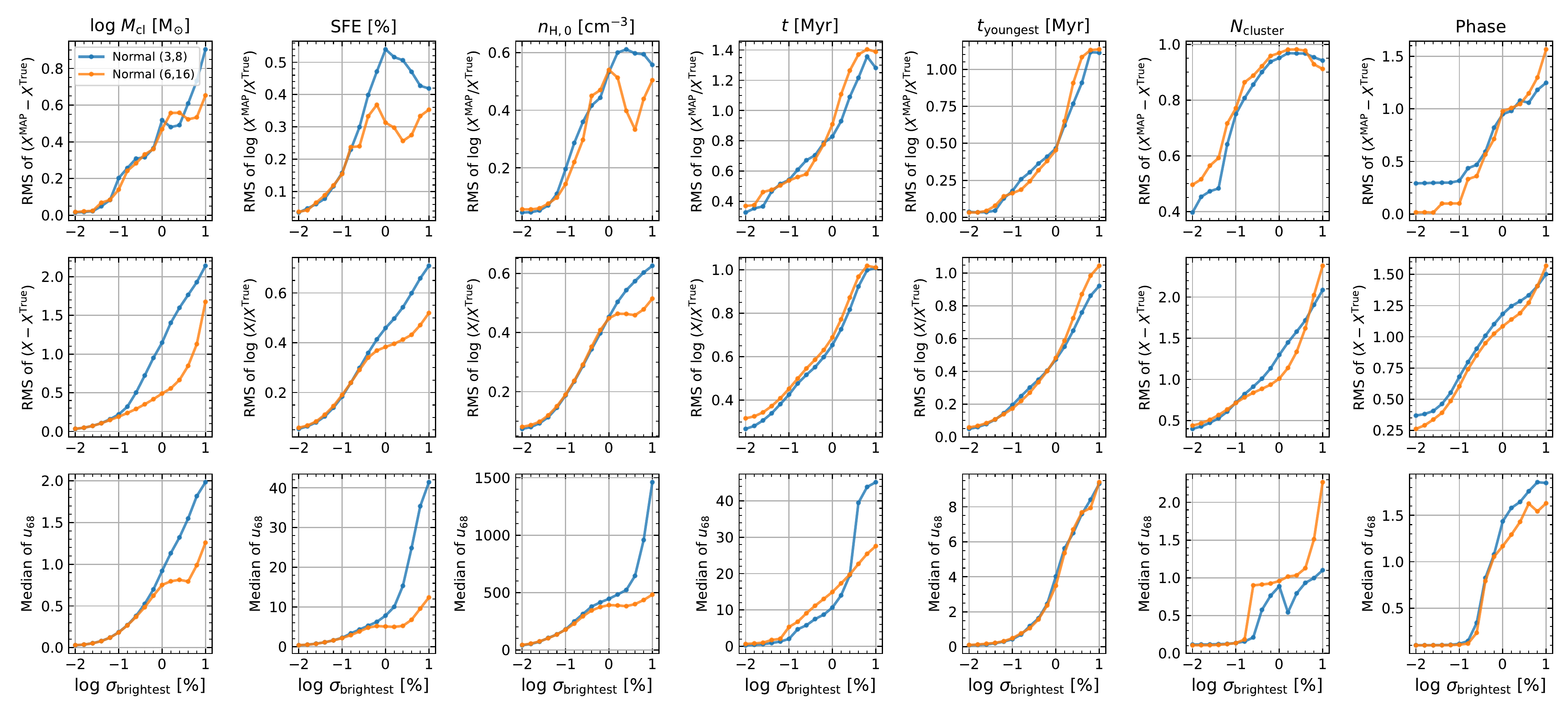}
    \caption{
    Comparison of two \normals\ with different numbers of blocks and layers of the internal sub-network. The orange line with 6 layers and 16 blocks is the same \normal\ used in the main contents (i.e., the red line in Figure~\ref{fig:rmse}).
     }\label{fig:normal_test}
\end{figure*}

Additionally, we investigate whether the performance of the \normal\ changes when increasing $N_\text{layer}$ and $N_\text{block}$. We only compare two networks using the setup of \citetalias{Kang+22} ($N_\text{layer}$=3, $N_\text{block}$=8) and the setup of the best \noise\ ($N_\text{layer}$=6, $N_\text{block}$=16). Figure~\ref{fig:normal_test} shows that the overall change of the \normal\ is different to the case of the \noise. 
The performance gap between the two networks becomes noticeable for \mcl, SFE and \hdenini, when the error is large. On the other hand, there is no significant change in the \ncl\ and $t$ prediction and sometimes the deeper network performs slightly worse than the other. In conclusion, we choose the deeper \normal\ (orange curve) and keep the network setup the same as the \noise\ for the main paper because the performance of the deeper \normal\ is not significantly worse, but rather exhibits some improvement for some parameters.

\bsp	
\label{lastpage}
\end{document}